\documentclass[conference]{IEEEtran}
\IEEEoverridecommandlockouts
\usepackage[utf8]{inputenc}
\usepackage{cite}
\usepackage{amsmath,amssymb,amsfonts}
\usepackage{mathrsfs}
\usepackage{algorithmic}
\usepackage{tikz}
\usetikzlibrary{quantikz2}
\usepackage{graphicx}
\usepackage{float}
\graphicspath{{figures/}}
\usepackage{textcomp}
\usepackage{orcidlink}
\usepackage{booktabs}
\usepackage{array}
\usepackage{tabularx}
\newcolumntype{Y}{>{\centering\arraybackslash}X}
\newcolumntype{L}{>{\raggedright\arraybackslash}X}
\newcommand{\thh}[1]{\shortstack[c]{#1}}
\usepackage{url}
\usepackage{xcolor}
\usepackage{colortbl}
\definecolor{qnowcolor}{HTML}{364483}
\usepackage{hyperref}
\hypersetup{
  colorlinks=true,
  allcolors=qnowcolor
}

\makeatletter
\newcommand{\linebreakand}{%
  \end{@IEEEauthorhalign}
  \hfill\mbox{}\par
  \mbox{}\hfill\begin{@IEEEauthorhalign}
}
\makeatother

\begin{document}

\title{Quantum-Enhanced QSAR for DYRK1A: Evaluating Complementary and Utility QPU-Derived Molecular Representations}

\author{
    \IEEEauthorblockN{Alejandro Giraldo \orcidlink{0009-0008-9826-0703}}
    \IEEEauthorblockA{
        \href{https://www.globant.com/}{$\mathtt{Globant}$}, \href{https://qnow.tech/}{$\mathtt{QNOW \; Technologies}$}\\
        Buenos Aires, Argentina\\
        \href{mailto:alejandro.gl@globant.com}{alejandro.gl@globant.com}
    }
    \and
    \IEEEauthorblockN{Daniel Ruiz \orcidlink{0009-0007-6976-1755}}
    \IEEEauthorblockA{
        \href{https://qnow.tech/}{$\mathtt{QNOW \; Technologies}$}\\
        Delaware, USA\\
        \href{mailto:daniel@qnow.tech}{daniel@qnow.tech}
    }
    \and
    \IEEEauthorblockN{Mariano Caruso \orcidlink{0000-0002-7455-1193}}
    \IEEEauthorblockA{
        \href{https://www.ugr.es/}{$\mathtt{UGR}$}, \href{https://www.unir.net/}{$\mathtt{UNIR}$}, \href{https://www.fidesol.org/}{$\mathtt{FIDESOL}$}\\
        Granada, Spain\\
        \href{mailto:mcaruso@fidesol.org}{mcaruso@fidesol.org}
    }
    \linebreakand
    \IEEEauthorblockN{Hamed Mohammadbagherpoor}
    \IEEEauthorblockA{
        \href{https://www.ncsu.edu/}{$\mathtt{IBM \ T. J. \ Watson \ Research \ Center}$}\\
        Yorktown Heights, NY, USA\\
        \href{mailto:hmohamm2@alumni.ncsu.edu}{hamed.mohammadbagherpoor@ibm.com}
    }
    \and
    \IEEEauthorblockN{Vladyslav Bohun \orcidlink{0000-0001-6937-1035}}
    \IEEEauthorblockA{
        \href{https://haiqu.ai/}{$\mathtt{Haiqu, Inc.}$}\\
        New York, USA\\
        \href{mailto:vladbohun@haiqu.ai}{vladbohun@haiqu.ai}
    }
}

\maketitle

\begin{abstract}
In drug discovery, hit identification selects a limited set of compounds for experimental evaluation from a larger collection of candidates. Because laboratory screening capacity is limited, that choice determines which chemical series continue into further investigation. A hit that is not assayed does not enter confirmation or the analog series, and that chemical series is not opened.
For the DYRK1A kinase, the same test molecules were scored from a tabular representation of molecular descriptors and from a vector measured on a quantum processor, with the same RBF-SVM. The question is whether the two models propose the same set for experimental evaluation: they propose different compounds. The model on the vector measured on the quantum processor includes compounds that the tabular representation leaves out, and among them are already-labeled actives. The tabular representation includes actives that the vector measured on the quantum processor leaves out.
What is reported is that set: who enters, with what label, and whether published chemistry already mentions it. No formal quantum advantage is claimed, nor a mechanism for why the two vectors separate differently. Those compounds fall within the neighborhood that both representations already cover; what changes is which model marks them active. From that follows the cost of predictive disagreement: each model would send a different chemical series to experimental evaluation.
\end{abstract}


\section{Introduction}
\label{sec:introduction}

In the drug discovery funnel, many structures are proposed and only a handful are chosen for assay; only a small fraction of the screened compounds advances through that funnel. The entire molecule library cannot be assayed: every early choice spends a campaign cycle, budget, and plate time, the physical laboratory space where each sample is deposited and measured, and leaves out other molecules that will never reach the assay. If a potentially active compound is not selected for experimental evaluation, that structure-activity hypothesis is never measured, the corresponding chemical series may never be explored in the laboratory, and the rest of the funnel works on different chemistry.

That choice of who gets assayed is hit identification. A hit is a molecule that, in a primary assay, gives a signal of activity against the target of study. It is the laboratory's first yes: this compound hit the target hard enough to deserve confirmation and, if it holds, a series of analog compounds. Every extra compound we assay spends a well, a compartment of the plate; a hit we do not assay does not enter confirmation or the analog series, and that chemical series is not opened.

\textit{Virtual screening} sieves libraries of small molecules and proposes candidate \textit{hits} for the \textit{in vitro} or \textit{in vivo} assays that come after: it reduces how many candidates reach that stage and orders their selection. The quantitative structure-activity relationship (QSAR) is one of the main approaches within \textit{virtual screening} \cite{neves2018}.

QSAR builds models that relate molecular structure, represented as a vector of descriptors, to a property previously measured on a set of training compounds. Once validated, the model can be used to predict that property in molecules that have not been experimentally assayed and, from those predictions, prioritize candidates for evaluation \cite{cherkasov2014}.

The QSAR model is built without needing the experimental assay of the proposed \textit{hits}; that assay comes later, as validation of its predictions. The prediction orients which compounds are taken to assay.

In this work, the assay is for dual-specificity tyrosine-phosphorylation-regulated kinase 1A (DYRK1A). The gene sits in the Down syndrome critical region and has been implicated in Tau and amyloid pathology associated with early-onset Alzheimer's disease \cite{coutadeur2015}. That literature explains why the assay exists, not what is claimed about a model.

Classical QSAR works on tables of molecular descriptors. This study investigates whether a quantum processor can give another molecular representation for QSAR modeling. In \textit{Quantum QSAR} we explored this with quantum support vector machines: whether a Hilbert space and its separability hold up on a different endpoint \cite{giraldo2025qqsar}. In \textit{QSAR} (Q²SAR), we moved to multiple quantum kernel learning, already on the classification of DYRK1A inhibitors, and that result was read through the area under the curve \cite{giraldo2025q2sar}. Here the technique is an \textit{entangled manifold embedding} (EME) with a local projected readout, and the comparison is the set that each representation proposes for experimental evaluation. That readout is the vector the processor returns for each molecule; what decides the label is still a classical classifier. This study evaluates the prediction differences that arise from the molecular representations, and it does not test a general quantum advantage in QSAR.

Supervised classification has been run with quantum-enhanced vectors on a processor \cite{havlicek2019}. Two models can both land above chance, even with similar confusion counts, and still propose different lists of candidate compounds. That disagreement leaves hits out of the plate and chemical series unopened. Predictive discrepancy names exactly that: classifiers with similar metrics and different patterns; the cost sits in the instances where they disagree \cite{renard2024}. Here those instances are wells and chemical series: marking active a compound that is not costs a well; marking inactive a hit leaves that series out. Before reaching the well, those compounds still go through a selection filter, inspection that builds the final assay list.

Ranking metrics, such as the area under the curve, precision, and other classification metrics, summarize the ordering of compounds in the test set according to the model's predicted activity score. The proposed candidate list depends on the classification threshold applied to the model's decision function. That metric can be won, even tying the hit count with another model, and still not agree on the label. QSAR validation practice reports that full menu of metrics and reads as the activity prediction whichever method comes out winning on any of them \cite{neves2018}. Modeling only serves its purpose if it addresses the problem actually posed here: which compounds move to experimental evaluation, not which metric comes out winning. That is, two models can share the metric and not share the list of compounds.

Our question for this study is: do the classical descriptor representation and the representation obtained on the quantum processor propose different compounds for experimental evaluation, on the same DYRK1A test set? We look at who ends up marked active under each representation. A ranking figure does not answer it, a disagreement does.

\section{Fundamentals}

The experiment classifies with a kernel SVM. The state on the processor is prepared by the \textit{entangled manifold embedding}, and the local readout of the X, Y, and Z observables returns the 213-expectation vector on which that classifier assigns the label.

\subsection{Kernel-Based Classification}
The formulation of the Support Vector Machine ($\mathtt{SVM}$) centers on maximizing the separation margin between classes within an induced feature space \cite{cortes1995support}. Given a training dataset $\mathscr{D} = \{(x_i, y_i)\}_{i=1}^M$, where $x_i \in \mathbb{R}^N$ represents the vector of molecular descriptors and $y_i \in \{-1, +1\}$ denotes the biological activity label, the soft-margin primal optimization problem is formulated as:
\begin{equation}
    \min_{w, b, \xi} \left( \frac{1}{2} \|w\|^2 + C \sum_{i=1}^M \xi_i \right)
\end{equation}
subject to the inequality constraints:
\begin{equation}
    y_i \left( \langle w, \phi(x_i) \rangle_{\mathscr{H}} + b \right) \ge 1 - \xi_i, \quad \xi_i \ge 0, \quad \forall i \in \{1, \dots, M\}
\end{equation}
where $w$ denotes the normal vector to the separating hyperplane in the Hilbert space $\mathscr{H}$, $b \in \mathbb{R}$ is the bias term, $C > 0$ is the regularization constant penalizing classification errors, and $\xi_i$ represent the slack variables.

Through the Lagrange dual formulation parameterized by the multipliers $\alpha_i$, the quadratic optimization problem is reformulated depending exclusively on inner products:
\begin{equation}
    \max_{\alpha} \sum_{i=1}^M \alpha_i - \frac{1}{2} \sum_{i=1}^M \sum_{j=1}^M \alpha_i \alpha_j y_i y_j K(x_i, x_j)
\end{equation}
subject to $\sum_{i=1}^M \alpha_i y_i = 0$ and $0 \le \alpha_i \le C$ for all $i \in \{1, \dots, M\}$.

The kernel function $K(x_i, x_j) = \langle \phi(x_i), \phi(x_j) \rangle_{\mathscr{H}}$ evaluates the inner product without explicitly computing the non-linear mapping $\phi(x)$. In a Quantum Support Vector Machine ($\mathtt{QSVM}$), that mapping is instantiated with a parameterized quantum circuit $U(x)$, which encodes the input vector $x$ into a quantum state $|\psi(x)\rangle = U(x) |0\rangle^{\otimes n}$ defined over a Hilbert space of dimension $2^n$ \cite{RebentrostEtAl2014PRL, havlicek2019}.

\subsection{Quantum Data Encoding via Entangled Manifold Embedding}
In noisy intermediate-scale quantum ($\mathtt{NISQ}$) devices, conventional data encoding strategies impose severe constraints on circuit depth or the required qubit count. To circumvent this limitation, the architecture uses the \textit{entangled manifold embedding} ($\mathtt{EME}$) \cite{schuld2019quantum,lloyd2020quantum, havlicek2019, haiqueme}.

The $\mathtt{EME}$ protocol defines an injective map $\phi_{\mathtt{EME}}: \mathbb{R}^N \to \mathscr{M}_{\mathscr{H}}$ from the classical descriptor space into a submanifold $\mathscr{M}_{\mathscr{H}} \subset \mathscr{H}$, characterized by a controlled entanglement density $D$ \cite{yyvr-dtsb}. When $D = 1$, the mapping reduces to unentangled product states, operating analogously to standard angle encoding. For $D > 1$, $\mathtt{EME}$ induces entangled states that encode high-dimensional vectors ($N$ components) using a reduced number of qubits ($n \ll N$), balancing entanglement entropy against circuit depth \cite{yyvr-dtsb}.

The synthesis of the state preparation unitaries $U_{\mathtt{EME}}(x)$ is approximated using matrix product state ($\mathtt{MPS}$) representations \cite{yyvr-dtsb}. This tensor decomposition bounds the bond dimension $\chi$, controllably truncating the entanglement spectrum, and so makes both the classical simulation and the compilation on the quantum processing unit ($\mathtt{QPU}$) tractable \cite{vidal2003, yyvr-dtsb}.

Following the initial embedding $U_{\mathtt{EME}}(x)$, an ansatz $V$ with a fixed entangling topology is applied. In the ideal unitary regime, this process yields the pure state
\begin{equation}
    |\psi(x)\rangle = V U_{\mathtt{EME}}(x) |0\rangle^{\otimes n}
\end{equation}
whose fixed parameters act as a frozen quantum feature map for the subsequent classical classification.

\subsection{Projected Quantum Kernels}
The application of fidelity quantum kernels ($\mathtt{FQK}$), defined by the global state overlap $K_{\mathtt{FQK}}(x_i, x_j) = |\langle \psi(x_i) | \psi(x_j) \rangle|^2$, shows critical vulnerabilities as the qubit register size $n$ grows. In high-dimensional Hilbert spaces, arbitrarily generated states tend asymptotically toward mutual orthogonality, and a concentration of measure appears in which $K_{\mathtt{FQK}}(x_i, x_j) \to \delta_{ij}$ exponentially. This gives rise to barren plateaus in the kernel landscape and degrades generalization capability \cite{HuangEtAl2021NatComm, Thanasilp2024}.

To circumvent that exponential concentration and give the model robustness against hardware noise, the projected quantum kernel ($\mathtt{PQK}$) framework is incorporated \cite{HuangEtAl2021NatComm, Schnabel_2025, D_Amore_2026}. Rather than evaluating the global state overlap, $\mathtt{PQK}$ projects the quantum system onto a classical space by extracting the one-body reduced density matrices (1-$\mathtt{RDM}$) \cite{HuangEtAl2021NatComm}.

When run on noisy quantum hardware, the resulting physical state is generally described by a density matrix (mixed state) $\rho(x) \in \mathcal{S}(\mathscr{H})$, subject to quantum noise channels $\mathscr{E}$ such that $\rho(x) = \mathscr{E}\left( |\psi(x)\rangle \langle \psi(x)| \right)$, where $\mathcal{S}(\mathscr{H})$ denotes the set of density operators acting on $\mathscr{H}$. In the noiseless limit, this description reduces to the pure-state projector $\rho(x) = |\psi(x)\rangle \langle \psi(x)|$.

For each qubit $k \in \{1, \dots, n\}$, the local reduced density matrix is obtained by tracing out the remaining subsystems: $\rho_k(x) = \mathrm{Tr}_{\neg k}\left[ \rho(x) \right]$. That projection is characterized by the expectation values of the Pauli observable set $\mathscr{P} = \{X, Y, Z\}$, and forms a classical vector $g(x) \in \mathbb{R}^{3n}$ whose components are $\mathrm{Tr}\left[ P \rho_k(x) \right]$.

The similarity between two samples $x_i$ and $x_j$ is modeled by evaluating those readout vectors within a classical Gaussian (RBF) kernel \cite{HuangEtAl2021NatComm}:
\begin{equation}
\begin{split}
    K_{\mathtt{PQK}}(x_i, x_j) &= \exp \left( -\gamma \| g(x_i) - g(x_j) \|^2 \right) \\
    &= \exp \Bigg[ -\gamma \sum_{k=1}^n \sum_{P \in \mathscr{P} } \Big( \mathrm{Tr}[P \rho_k(x_i)] \\
    &\qquad\qquad\qquad\qquad\quad - \mathrm{Tr}[P \rho_k(x_j)] \Big)^2 \Bigg]
\end{split}
\end{equation}
where $\gamma > 0$ is the kernel scale hyperparameter. Since the sum of squared Pauli expectation differences is proportional to the Frobenius distance $\|\rho_k(x_i) - \rho_k(x_j)\|_F^2$, this projective representation preserves non-linear local correlations, mitigates the exponential concentration, and maximizes reliability on $\mathtt{NISQ}$-era processors \cite{HuangEtAl2021NatComm, Schnabel_2025}.

\subsection{Quantum Multiple Kernel Learning}
Because individual chemical representations often carry geometric biases and partial coverage of the space, the framework integrates multiple kernel learning ($\mathtt{MKL}$) \cite{gonen2011multiple} through its quantum extension ($\mathtt{QMKL}$) \cite{giraldo2025qqsar}. Instead of a single kernel matrix, $\mathtt{QMKL}$ optimizes a convex linear combination over a dictionary of $m$ base kernels $\{K_l\}_{l=1}^m$, bringing together classical kernels and projected quantum kernels:
\begin{equation}
    K_{\mathtt{MKL}}(x_i, x_j) = \sum_{l=1}^m w_l K_l(x_i, x_j), \quad w_l \ge 0, \quad \sum_{l=1}^m w_l = 1
\end{equation}
where the weights $w_l$ are determined during the classical optimization phase.

The QMKL of the prior work combines several kernels and was scored by the area under the curve. In this run the classifier receives the 1080-descriptor table, or the 213 measured expectations.

\section{Experiments}
\label{sec:methodology}

The design follows the Introduction's question: when the same DYRK1A test molecules are represented with the classical descriptor table or with the vector measured on the quantum processor, do the two representations produce different candidate lists for experimental evaluation?

To answer it, the two models were trained on the same training set and evaluated on the same test set, with the same rule for assigning a compound to the list of compounds predicted as active.

The main comparison used a single training and test split and a single classifier protocol. Other stratified partitions were evaluated afterward, on the same already-computed molecular representations, including the already-measured QPU vectors, so no new QPU measurements were needed.

We evaluated the protocol with three molecular representations. The first was a Mordred descriptor representation, of dimension 1080. The other two were generated with the same 71-qubit \textit{entangled manifold embedding} (EME) circuit from Haiqu, evaluated once on a matrix product state (MPS) simulator and once on the IBM Quantum processor, with the \texttt{ibm\_boston} backend.

The matrix product state (MPS) method represents the quantum state as a chain of local tensors and truncates the bond dimension. For states whose entanglement is sufficiently limited, that representation can require substantially less memory than the explicit vector of \(2^n\) amplitudes \cite{vidal2003}, which is what makes it possible to simulate the 71-qubit circuits used here.

The main comparison uses an RBF-SVM trained separately on the classical descriptor representation and on the representation obtained on the QPU. The same vectors were also evaluated with four other classifier families: logistic regression, random forest, \textit{gradient boosting}, and \(k\)-nearest neighbors. The goal was to see whether the observed disagreement depended on the classifier family.

That reading, whether another family would flag the same molecules, is declared later.

The matrix product state simulator evaluates the same circuit as the QPU. The list comparison is the table's RBF-SVM against the RBF-SVM of the vector measured on the QPU.

A model proposes a compound when the RBF-SVM's decision function is positive: the signed distance to the hyperplane in kernel space. That rule is the estimator's default label; no new threshold was chosen on the test set.


\subsection{Data and molecular descriptors}
\label{sec:data-split}

To compare the two lists, both models saw the same molecules. The set is 354 compounds with a binary DYRK1A activity label (active/inactive), the same one used in \textit{QSAR} (Q²SAR) \cite{giraldo2025q2sar}. The binary label comes from an earlier curation, target CHEMBL2292 and curation of the ProtoQSAR dataset, recorded as active if the IC50 was less than or equal to \(1\,\mu\mathrm{M}\). This run starts from that label \cite{zdrazil2024}.

Each molecule entered as a SMILES (\textit{Simplified Molecular Input Line Entry System}): a string that encodes the structure. The IC50 is the concentration, in micromoles per liter, that leaves the enzyme at half its activity; that cutoff fixes who counts as a hit in the original assay. This run did not reapply the cutoff.

We fixed a split of 264 training compounds and 90 test compounds. Active-class rates were 0.561 in train and 0.556 in test: the task is nearly balanced, so \textit{accuracy} is a reasonable measure here, but it is not this question's criterion. Molecular descriptors were computed by the Mordred software: 1080 columns, already standardized. We aligned train and test to those same columns. Structures were not recomputed and the table's dimensionality was not reduced. Those 90 are the test set of the main comparison.

\subsubsection{Metrics}
\label{sec:metrics}

The two representations were judged with the same QSAR classification protocol. The QPU vector went into the same figures as the table. An area under the curve can tie while the proposed molecules do not match.

For the active class we reported F1, precision, recall, and the Matthews correlation coefficient (MCC), plus \textit{accuracy} and \textit{balanced accuracy}. The rule for list membership is the one already declared: a positive decision function, the sign of the distance to the RBF-SVM hyperplane defined earlier.

Hyperparameter selection was based on the F1 of the active class, which balances precision and recall and does not incorporate true negatives. MCC was reported in addition because it incorporates all four entries of the confusion matrix \cite{chicco2020}. Recall and precision quantify, respectively, sensitivity to active compounds and the fraction of predicted actives that are correct; \textit{accuracy} and \textit{balanced accuracy} provide complementary measures of classification performance \cite{oecd2007}.

ROC-AUC and PR-AUC were computed from the model's continuous scores. They give threshold-independent measures of discrimination and ranking over the 90 test molecules.

The enrichment factor, BEDROC, the geometric mean, and the positive predictive value at plate size, the fraction of actives among the top-N ranked compounds when N is the size of an assay plate, are declared as a first step toward early recognition, the problem that Truchon and Bayly address for virtual screening \cite{truchon2007}, with no figure yet in this body. The Cooper statistic is used later, in the Results applicability domain \cite{cooper1979}.

\subsection{Quantum circuits}
\label{sec:eme-circuits}

A quantum processor can give another representation of the same molecule. We follow the projected quantum kernel scheme \cite{Schnabel_2025,D_Amore_2026}: the real classical data is loaded into a wavefunction and an ansatz acts on it before measurement.

Each entry in the set is loaded separately via Haiqu's \textit{entangled manifold embedding} (EME). The method defines a bijection between real vectors and a manifold of Hilbert space with fixed entanglement. That entanglement is controlled by the density parameter $D$. With $D=1$ the state is a product and matches the usual angle encoding. Larger values allow encoding more data in fewer qubits, and the circuit grows. That leaves a trade-off between the width and depth of the preparation circuits.

This run's table has 1080 descriptors, and the measured vector has 213 expectation values, three for each of the 71 qubits, with density $D=4$. Concatenating these two representations sums to 1293 components, and that 1293-dimensional vector is left for later: it is not the circuit's input. The state-preparation circuits are synthesized with MPS methods \cite{yyvr-dtsb}.

Each preparation is followed by a shared ansatz. It has a brick-wall structure: one layer on the even couplings (0-1, 2-3, …) and one layer on the odd couplings (1-2, 3-4, …). Each of those gates consists of two CX gates and a pair of RY and RZ rotations between each coupling. The ansatz ends with a layer of single-qubit RY and RZ gates. Figure~\ref{fig:kernel_circuit} is the shared ansatz that follows the EME block. The ansatz parameters were fitted once on a training set, with F1 and the SPSA optimizer, a stochastic-gradient method, and are shared across the experiments.
From each circuit, single-qubit expectation values in X, Y, and Z are collected. This yields $3\times71=213$ values, and those values are the set processed on the quantum processor.

\begin{figure}[H]
\centering
\resizebox{\columnwidth}{!}{%
\begin{quantikz}[row sep={0.9cm,between origins}, column sep=0.25cm]
\lstick{$q_1$}
  & \gate[6]{\mathrm{EME}}
  & \gate{R_y} & \gate{R_z} & \ctrl{1} & \gate{R_y} & \gate{R_z} & \ctrl{1}   
  & & & & & &                                                                  
  & \gate{R_y} & \gate{R_z} & \\
\lstick{$q_2$}
  &
  & \gate{R_y} & \gate{R_z} & \targ{} & \gate{R_y} & \gate{R_z} & \targ{}     
  & \gate{R_y} & \gate{R_z} & \ctrl{1} & \gate{R_y} & \gate{R_z} & \ctrl{1}   
  & \gate{R_y} & \gate{R_z} & \\
\lstick{$q_3$}
  &
  & \gate{R_y} & \gate{R_z} & \control{}\wire[d][1][shorten >=0.2cm]{q}
  & \gate{R_y} & \gate{R_z} & \control{}\wire[d][1][shorten >=0.2cm]{q}      
  & \gate{R_y} & \gate{R_z} & \targ{} & \gate{R_y} & \gate{R_z} & \targ{}     
  & \gate{R_y} & \gate{R_z} & \\
\wave &&&&&&&&&&&&&&&&& \\
\lstick{$q_{n-1}$}
  &
  & \gate{R_y} & \gate{R_z} & \targ{}\wire[u][1][shorten >=0.2cm]{q}
  & \gate{R_y} & \gate{R_z} & \targ{}\wire[u][1][shorten >=0.2cm]{q}         
  & \gate{R_y} & \gate{R_z} & \ctrl{1} & \gate{R_y} & \gate{R_z} & \ctrl{1}   
  & \gate{R_y} & \gate{R_z} & \\
\lstick{$q_n$}
  &
  & & & & & &                                                                  
  & \gate{R_y} & \gate{R_z} & \targ{} & \gate{R_y} & \gate{R_z} & \targ{}     
  & \gate{R_y} & \gate{R_z} &
\end{quantikz}

}
\caption{Shared ansatz after the EME block.}
\label{fig:kernel_circuit}
\parbox{\columnwidth}{\small Even and odd couplings alternate. The EME block loads each sample. The readout is single-qubit X, Y, and Z.\vspace{0.55em}}
\end{figure}

\subsection{Backends}
\label{sec:backends}

A circuit is not a vector until it is measured. The circuit fixes the state; the vector is the 213 expectation values that come out of measuring it. The same circuit was measured on a matrix product state simulator and on the QPU. The 213-component vectors were read from this run's cache. Shots, bond dimension, and mitigation fall outside this record.

\subsection{Classifier protocol}
\label{sec:classical-head}

With the 1080-descriptor table and the two 213-expectation vectors, the MPS simulator's and the QPU's, each classifier builds a list of compounds to assay under the same active rule. We fixed a protocol. Logistic regression, the RBF-SVM, and k-nearest neighbors were standardized with a scaler fit only on train. Each cross-validation fold fit that scaler on its own training rows. The random forest and \textit{gradient boosting} were trained without that layer.

Hyperparameters were selected with five-fold stratified cross-validation, restricted to the training set. For each classifier and each representation, the grid configuration that maximized the active-class F1 on the training folds was chosen; the selected model was then evaluated on the held-out test set.
That test set, 90 compounds not used to choose hyperparameters, was evaluated only once training and model selection were already finished.

\subsection{Disagreement set}
\label{sec:disagreement}

On the same 90 test compounds, an RBF-SVM was applied, separately, to the classical descriptor representation and to the representation obtained on the QPU. Each model assigned each compound to the predicted-active or predicted-inactive class, under the already-defined decision-function threshold. Disagreement occurs when the two models assign different classes to the same compound.

QPU-only actives have a positive decision function on the QPU and a negative one on the table. Classical-only actives are the reverse. For each compound in those sets we kept the laboratory label and the SMILES.

\subsection{Additional classifier families}
\label{sec:families}

On the same vectors and the same grid protocol, we compared the five families by representation: logistic regression, RBF-SVM, random forest, \textit{gradient boosting}, and \(k\)-nearest neighbors.

We evaluated four other classifier families to see whether the disagreement observed in the candidate lists also appeared with classifiers other than the RBF-SVM.

Within each family, the grid still chooses hyperparameters by the active-class F1 on train. The remaining figures are reported on test; they do not choose that grid.

Across families, two test readings are declared, the highest F1 and the highest ROC-AUC. The pair in question remains the table's RBF-SVM against the QPU's.
The table's RBF-SVM against the QPU's remains the pair in question, and its grid is chosen on training. Concatenation, stacking, and the Shapley split use, for each representation, the family with the highest ROC-AUC on this partition's test set. Because that choice uses the test set, those three readings remain exploratory for this partition.

\subsection{Split-robustness analysis}
\label{sec:split-robustness}

The 264/90 split is one. If the actives that only the QPU proposes, and that the laboratory also marks active, were an accident of that partition, they would not reappear in others. We repooled the 354 compounds with the already-measured QPU vectors and repeated the pair, the table's RBF-SVM against the QPU's RBF-SVM, on ten new stratified 264/90 partitions. For each one we recorded whether those molecules fall in test and who proposes them, and the ROC-AUC on each side. Those ten partitions are a modest sample. Frequency is declared as a count: how many times the molecule appears in test, and in how many of those the QPU alone proposes it. These ten repetitions are not enough to generalize beyond themselves; that is why a later extension repeated the same protocol over a hundred new partitions instead of ten, to give a conditional rate with a confidence interval instead of that raw count. Those numbers, outside the original sample, are reported in Results and Discussion.

That extension reuses the code and the ten original seeds unchanged. Before trusting the ninety new partitions, seeds 0 through 9 had to reproduce, figure for figure, the already-published table of the ten original partitions. That exact reproduction is the extension's acceptance condition; without it, the remaining ninety were not run.

Across those hundred partitions we searched, besides 8d and XMD14-124, for which other active molecules appeared as QPU-only more than once. That search is raw-frequency: it counts how many times each molecule falls into some partition's QPU-only set, without applying the rest of the protocol, applicability domain and decision margin, that was applied to 8d and XMD14-124. The catalog's most recurrent candidates are investigated in Results; the rest remain only catalogued, as candidates for future study.

\subsection{Applicability domain}
\label{sec:applicability}

A QSAR model only interpolates where training already covers the space. The applicability domain is the chemical, structural, or biological space where the model is reliable for prediction. OECD guidance calls for characterizing that space \cite{oecd2007}. Here we measure it as distance to the \(k\)-th train neighbor, in the already-scaled space of each ROC-AUC table's kNN: if a test compound sits farther than the threshold fixed on train, the prediction is extrapolation. That distance is Minkowski, with exponent \(p\): \(p = 1\) is Manhattan and \(p = 2\) is Euclidean; we use the \(p\) of that kNN.

The threshold is fixed only on train: the 95th percentile of each train point's distance to its \(k\)-th neighbor, excluding itself. A test compound is out if it exceeds that threshold. The percentile is declared in advance; it is not chosen by looking at the test set.

The kNNs that define the pair's space are: table, k=5 and Manhattan; QPU vector, k=11 and Euclidean. The concatenation's, 1293 columns, k=5 and Euclidean, is declared with that recipe later.

As a structural check, not a descriptor one, we measured the maximum Tanimoto similarity of each test compound against train. Tanimoto compares two binary fingerprints, the ratio of shared bits to bits set in either one. The fingerprints are Morgan, radius 2, 2048 bits: a map of atomic environments. The 0.70 cutoff was declared in advance. This check does not replace the descriptor domain; the kNN does not decide in fingerprint space.

In QSAR the applicability domain is a condition for model validity \cite{oecd2007, eriksson2003}. A compound \textit{inside} has neighbors in train: the label interpolates a piece of the space the classifier already saw. A compound \textit{outside} does not have them: the same mark, active or inactive, is extrapolation, and does not hold with the same strength. If a disagreement molecule falls outside that applicability space, that prediction lacks coverage: neither classifier backs it with neighbors already seen in train.

\subsection{Boundary inside the domain}
\label{sec:boundary}

A compound inside the domain can fall on either side of the decision surface. That surface defines the classifier's boundary: a positive decision function places the compound on the proposed list, and a negative one excludes it. Far from the boundary, the 0/1 label is stable. Close to the boundary, a small change can flip the classification, and two models can disagree only by the threshold. On the same split, and without new hardware submissions, we measured distance to the decision boundary, whether other classifiers in the same representation space gave the same prediction, and a cut of the already-trained surface.

The RBF-SVM's decision margin is the absolute value of the decision function: how far the molecule sits from the decision surface, the line that splits active from inactive in kernel space. If the classifier produces a probability, the corresponding distance is measured from 0.5. We kept that value for each test compound and compared it with the other 89. A small decision margin, relative to the rest of the compounds, indicates that molecule sits close to that model's surface \cite{watsondaniels2023}. Cutting that number to a 0/1 label can increase apparent disagreement between two models whose continuous scores are similar \cite{hsu2022}.

In each representation, table or QPU, several classifiers sit close to the best one on train: that predictive multiplicity, that near-equally-good models differ on a single prediction, is what the $\epsilon$-set quantifies here \cite{marx2020}. The $\epsilon$-set are those that, in train cross-validation, have an F1 at most 0.03 below that space's winner. That 0.03 is declared on train; it was not chosen by looking at the disagreement molecules. We counted how many of those classifiers propose the QPU-only actives that the laboratory also marks active. That asks whether the disagreement belongs to the RBF-SVM or to the whole space. The table-versus-QPU pair does not enter this count: they are two spaces.

The decision surface cannot be drawn in 1080 or in 213 dimensions, but two methods approximate its visualization. To see where the disagreement molecules fall relative to that already-trained surface, we cut a plane that passes through the train centroid and through those molecules, and read the sign of the original RBF-SVM on that plane \cite{somepalli2022}. That cut is a view of the pair's surface; it is not retrained. Additionally, a 2-D and 3-D PCA, with an SVM fit in that reduced space, represents another decision surface, and should not be read as a visualization of the original classifier in the full-dimensional representation.

A large decision margin and the same vote within the $\epsilon$-set support the list's label. A small margin, or a split vote, leave it hanging on the threshold. A QPU-only active far from the table's surface, with no vote in that table's $\epsilon$-set, falls outside with room to spare; pressed against that surface, or with a split vote, the 0/1 mark loosens. The same two readings are kept for the QPU.

\subsection{Combination of representations}
\label{sec:combination}

It remains to see whether combining the two representations changes which compounds are selected: whether compounds appear that each side, separately, leaves out. On the same split, and without generating new hardware submissions, we combined the two representations in two ways.

Concatenation joins, for each molecule, the table's 1080 columns and the QPU vector's 213 expectation values, yielding a 1293-component vector. The classifier is chosen with the same cross-validation and grid protocol already declared: five folds on train, active-class F1.

Stacking is a two-level ensemble: a final model learns to combine the scores, not the columns, of other already-trained classifiers. In this stacking, from each representation we take the classifier with the highest test ROC-AUC, the one declared in Families, and its out-of-fold score, each molecule's score when that classifier was trained without the fold containing it. On those two scores, table and QPU, the meta-model is a logistic regression; its coefficients are declared with this combination's ranking in Results.

Concatenation has its own domain: in the 1293 columns, Euclidean distance to the fifth train neighbor. The threshold is the 95th percentile of those train distances, the same criterion already declared. A test compound falls out if it exceeds it. Stacking combines two scores. Each molecule's domain is read in the spaces that already exist: whether it interpolates in the table, in the QPU vector, or in both. The logistic regression's score has no coverage threshold of its own.

The same concatenation, with the table representation alongside the MPS simulator's vector, serves as a control: same recipe, different backend. That control asks whether joining columns moves the list even without the QPU vector. The impact on the list and ranking, and the recovery of 8d and XMD14-124, are reported with that combination.

\subsection{Classifier ensemble control}
\label{sec:ensembles}

An ensemble combines the outputs of several already-trained classifiers so the combination generalizes better than any one alone. Concatenating or stacking scores mixes the table and the QPU vector. It is possible that combining classifiers, without combining those two vectors, moves the list just as much. To compare those two things we ensembled the five already-declared families, same split, no new submissions, same grid.

\textit{Voting} and \textit{stacking} were trained under three conditions. \textit{Voting} is a soft vote over those families; \textit{stacking} fits a logistic regression on their scores. The conditions are: the five on the table alone; the five on the QPU vector alone; and the ten together, each seeing only its own column block. For each one we kept the proposed list and the ranking, the same figures as in concatenation.

\textit{Voting} and \textit{stacking} combine classifiers that can be wrong on different compounds. Here they serve to determine whether combining classifier outputs, within one representation or across both, produces an effect similar to mixing the representations themselves. None of the three conditions was selected from the test set's results, and the Results section does not use these ensembles to support its main conclusions.

\subsection{Attribution of the two representations via Shapley values}
\label{sec:attribution}

Shapley values split the outcome of a cooperative game uniquely and additively among the players who produce it; here the game is concatenation and the players are the two column blocks, table and QPU. Concatenating the table and the QPU vector moves that list and the ranking. What remains is how much of that joint figure comes from each representation. The classical block is the table's 1080 columns; the QPU block, the 213 expectation values.

Four coalitions were measured: no model, chance; the table alone; the QPU vector alone; and both together, the concatenation. Each coalition's value is the test ROC-AUC. The Shapley split divides that joint figure between the two blocks with a closed formula: each receives the average of what it adds when it enters alone and when it enters with the other.

On the scores of those two classifiers, the one with the highest ROC-AUC for each representation, we measured correlation and disagreement \cite{kuncheva2003}: whether they score alike and on how many of the 90 test molecules they disagree.

In the already-defined two-input stacking model, SHAP attributes each prediction to the two input scores, and quantifies how much the table's score and the QPU's score contribute toward each molecule's predicted label. With only two inputs there are just two possible orderings to combine them, so that exact split is computed directly, without approximating it by sampling as a model with more inputs requires; it is the same marginal-contribution split, now applied per molecule instead of over the joint ROC-AUC.

In the table representation we measured the mutual information between each descriptor and the class label: how much information each descriptor carries about whether a compound is active or inactive. The same analysis was run on each component of the QPU vector.

\subsection{Noise characterization and calibration}

The QPU vector and the table vector can disagree either because the quantum embedding captures something the table does not see, or because hardware noise shifts the score without any real signal underneath. Before we can tell those two explanations apart for the six QPU-only molecules, we need to characterize how large that noise is and what shape it has. The vector measured on the QPU and the vector from the matrix product state simulator share the same circuit; only whether that circuit was evaluated on hardware or noise-free changes. The component-by-component difference between the two, on the 264 training molecules, is an already-observed sample of what shot noise, hardware noise, and error mitigation do to this specific preparation. That difference describes what was already measured between the two cached vectors, an empirical record, not a simulation of the backend's error physics.

On that difference we compute the mean and standard deviation of each of the 213 components, and the correlation between pairs of components. A strong correlation between neighboring components matters for choosing the noise model: an independent per-component noise model would understate a correlated joint shift between nearby qubits on the chip.

On the six QPU-only molecules, already identified in the Disagreement set, we start from each one's clean MPS-simulator vector and add synthetic noise calibrated from the already-characterized difference, before scoring it with the already-fitted QPU RBF-SVM; that classifier is not retrained. Two noise models are used: one resamples whole rows, already observed, from the training difference, and so preserves the correlation between components; the other samples each component separately from a normal with the same mean and standard deviation, deliberately discarding that correlation, as a check of whether the conclusion depends on it. The two readings per molecule are the fraction of repetitions in which the synthetic score crosses to the active side, and the percentile occupied by the real jump, the one observed between the clean and the measured vector, within that same simulated-noise distribution.

As a contrast control, the same procedure was run on the five test molecules with the most negative decision function on the table, the ones where the table and the QPU already agree on marking inactive with the largest margin. If synthetic noise crossed those five to the active side at the same rate as the six QPU-only molecules, the procedure would distinguish nothing; that is the baseline for reading the six molecules' crossing rate.

\subsection{Statistical significance}

The table wins the ranking and the QPU-only and classical-only compositions differ, but those figures come from just 90 test molecules. Before reading those differences as a real effect, we need to know whether they are defensible at that sample size or whether they fall within the expected sampling noise. On the same 90 test molecules, with the same decision function of the table's RBF-SVM and the QPU's RBF-SVM already used throughout this body, a paired \textit{bootstrap} resamples the test indices with replacement and applies the same resampling to both score vectors: this way the resampling always compares the same molecules on both sides, not two independent samples. From that resampling we report the ROC-AUC interval for each representation and for their difference, and what fraction of the repetitions flips which representation wins that ranking.

On the two ROC curves, built from the same labels, the DeLong test, the analytic alternative to bootstrapping for comparing two correlated ROC-AUCs, gives a closed form for that difference's p-value, without the variance that resampling introduces.

The active/inactive split of the QPU-only set against the classical-only set, already counted in the Disagreement set, is also compared with Fisher's exact test on that 2x2 table, and each rate's interval is reported with the Wilson method, which gives a reliable confidence interval for a proportion even with small samples. Those two counts, six and nine compounds, declare low statistical power in advance; the figure is reported with that caveat, not omitted.

\section{Results}
\label{sec:results}

\subsection{The two RBF-SVMs do not propose the same 90}
\label{sec:results-pair}

The pair is the table's RBF-SVM against the QPU's, same rule, same 90 test molecules. Table~\ref{tab:rbf-pair} reports this pair's test metrics over the same 90 molecules. Both rank above chance; the table wins that comparison (ROC-AUC 0.8790 versus 0.8255). The lists do not match: the QPU flags six compounds the table leaves out, two active and four inactive.

\begin{table}[H]
\centering
\caption{Test metrics for the RBF-SVM pair.}
\label{tab:rbf-pair}
\parbox{\columnwidth}{\small Both rows use the same 90 test molecules and the active class. F1 balances precision and recall at the decision-function threshold. ROC-AUC ranks the continuous scores. fp, fn, and tp are false positives, false negatives, and true positives.\vspace{0.55em}}
\footnotesize
\setlength{\tabcolsep}{4pt}
\begin{tabular*}{\columnwidth}{@{\extracolsep{\fill}}lccccc@{}}
\toprule
Representation & F1 & ROC-AUC & fp & fn & tp \\
\midrule
classical 1080-D & 0.8679 & 0.8790 & 10 & 4 & 46 \\
QPU            & 0.8155 & 0.8255 & 11 & 8 & 42 \\
\bottomrule
\end{tabular*}
\end{table}

On both sides the best test F1 is the RBF-SVM: table, tn 30, fp 10, fn 4, tp 46; QPU, tn 29, fp 11, fn 8, tp 42. The highest ROC-AUC among the five families is kNN: 0.8908 on the table, 0.8583 on the QPU. The pair in question remains RBF-SVM against RBF-SVM, the same family on both vectors.

The RBF-SVM's ROC-AUC on MPS is 0.8895; the same classifier drops to 0.8255 on the QPU (\(\Delta -0.064\)), and the gap between those two figures has no isolated cause. kNN barely moves (0.8560 \(\rightarrow\) 0.8583). Table~\ref{tab:five-heads} gives the test F1 and ROC-AUC of the five families on the descriptor table and on the QPU.

\begin{table}[H]
\centering
\caption{Test F1 and ROC-AUC by family.}
\label{tab:five-heads}
\parbox{\columnwidth}{\small Each row is one classifier family on the same 90. The left pair of columns is the 1080-descriptor table. The right pair is the QPU vector. F1 is the active class at the threshold. ROC-AUC ranks the continuous scores.\vspace{0.55em}}
\footnotesize
\setlength{\tabcolsep}{4pt}
\begin{tabular*}{\columnwidth}{@{\extracolsep{\fill}}lcccc@{}}
\toprule
& \multicolumn{2}{c}{table 1080-D} & \multicolumn{2}{c}{QPU} \\
\cmidrule(lr){2-3} \cmidrule(lr){4-5}
Family & F1 & ROC-AUC & F1 & ROC-AUC \\
\midrule
logistic regression & 0.8381 & 0.8360 & 0.7723 & 0.7920 \\
RBF-SVM             & 0.8679 & 0.8790 & 0.8155 & 0.8255 \\
random forest    & 0.8462 & 0.8798 & 0.7890 & 0.8245 \\
\textit{gradient boosting} & 0.8155 & 0.8595 & 0.7059 & 0.7275 \\
\(k\) nearest neighbors       & 0.8491 & 0.8908 & 0.7931 & 0.8583 \\
\bottomrule
\end{tabular*}
\end{table}

\subsection{Six QPU-only, nine classical-only}
\label{sec:results-counts}

\begin{figure}[H]
\centering
\includegraphics[width=0.72\columnwidth]{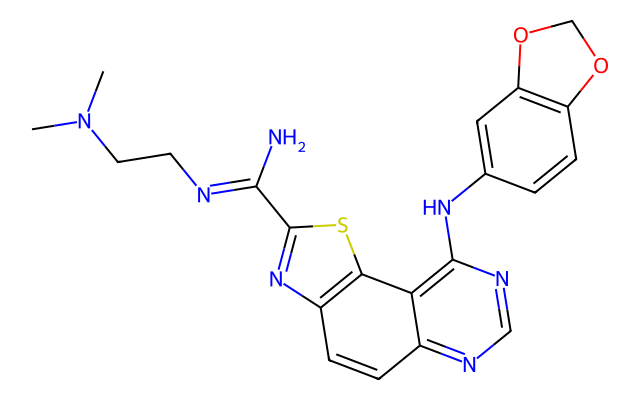}
\caption{8d, flagged only by the QPU.}
\label{fig:mol-8d}
\parbox{\columnwidth}{\small The descriptor RBF-SVM leaves it out. DYRK1A analogue (compound 33 of US~9{,}446{,}044). PubChem CID~71529733; ChEMBL CHEMBL2386756~\cite{foucourt2014a,leblond2016}.}

\vspace{0.6\baselineskip}
\includegraphics[width=0.72\columnwidth]{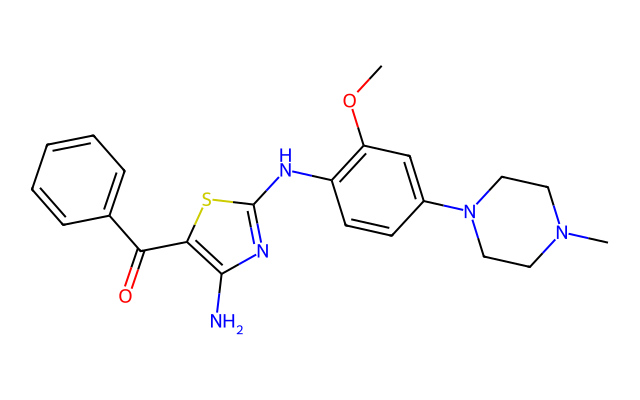}
\caption{XMD14-124, flagged only by the QPU.}
\label{fig:mol-xmd}
\parbox{\columnwidth}{\small The descriptor RBF-SVM leaves it out. A crystallographic probe, a molecule used to solve the protein's 3-D structure by X-ray crystallography rather than as a therapeutic candidate, ligand of PDB~6EIQ (code B6Z). PubChem CID~134812699; ChEMBL CHEMBL4171089~\cite{czarna2018}.}
\end{figure}

The table's SVM flags nine compounds that the QPU's leaves out; six of those nine are labeled active. Paying four extra inactive wells recovers two labeled hits the table's list would not have proposed. The table's list still proposes six labeled actives the QPU misses. The union of the two lists is a plate policy: instead of choosing between the two lists, every position either one proposes gets assayed. Table~\ref{tab:disagreement} counts the four list-membership cuts over those 90, and how many of each set are labeled active. The two labeled actives in that set are shown in Figures~\ref{fig:mol-8d} and~\ref{fig:mol-xmd}. The QPU's sign is positive and the descriptors' is negative. Table~\ref{tab:disagreement-lists} names the fifteen disagreement molecules, with index, label, both decision functions, and the SMILES.

\begin{table}[H]
\centering
\caption{List membership, same 90.}
\label{tab:disagreement}
\parbox{\columnwidth}{\small Each row is a slice of those 90. \(n\) is the number of molecules in the slice. Actives and inactives are the laboratory label within it. QPU-only and classical-only are the compounds one list proposes and the other leaves out. Both lists and neither list complete the partition.\vspace{0.55em}}
\footnotesize
\setlength{\tabcolsep}{4pt}
\begin{tabular*}{\columnwidth}{@{\extracolsep{\fill}}lccc@{}}
\toprule
Set & \(n\) & Active & Inactive \\
\midrule
QPU-only      &  6 &  2 &  4 \\
classical-only  &  9 &  6 &  3 \\
both lists  & 47 & 40 &  7 \\
neither list & 28 &  2 & 26 \\
\bottomrule
\end{tabular*}
\end{table}

\subsection{Two labeled actives are in the QPU-only set}
\label{sec:results-hits}

Neither of the two recovered actives is an approved drug for clinical use. One of them, 8d, is protected by a composition-of-matter patent, but a patent does not imply regulatory approval: it protects the invention, it does not certify that it has passed clinical trials. Identity is by SMILES against published DYRK1A chemistry.

\begin{table}[t]
\centering
\caption{Disagreement molecules.}
\label{tab:disagreement-lists}
\parbox{\columnwidth}{\footnotesize Table~\ref{tab:disagreement-lists} lists the disagreement molecules. \(f\) is the signed decision function; a positive value is that RBF-SVM's active side. Index is the compound id. The set column names the exclusive list. Label is the laboratory class. \(f\) table and \(f\) QPU are the two decision functions. The SMILES column keeps the first 16 characters and the last 8. Among the compounds the QPU lists and the descriptor table leaves out, the green row is labeled active and the red row, inactive. The SMILES of indices 62 and 87 link to the PubChem entry. 8d is index 87. XMD14-124 is index 62.\vspace{0.4em}}
\footnotesize
\setlength{\tabcolsep}{2pt}
\setlength{\aboverulesep}{0pt}
\setlength{\belowrulesep}{0pt}
\setlength{\extrarowheight}{1.2pt}
\begin{tabular*}{\columnwidth}{@{\extracolsep{\fill}}cllcc>{\raggedright\arraybackslash}p{0.36\columnwidth}@{}}
\toprule
Index & Set & Label & \(f\) table & \(f\) QPU & SMILES \\
\midrule
\rowcolor{red!12}
24 & QPU-only & inactive & \(-0.186\) & \(0.027\) & \texttt{Cc1ccc(C(=O)Nc2n}\newline\texttt{{\ldots}4)n2)cc1} \\
\rowcolor{red!12}
54 & QPU-only & inactive & \(-0.202\) & \(0.286\) & \texttt{Nc1ccc(N)c2c1C(=}\newline\texttt{{\ldots}ccc1C2=O} \\
\rowcolor{red!12}
59 & QPU-only & inactive & \(-0.069\) & \(0.410\) & \texttt{c1csc(CNc2ncnc3c}\newline\texttt{{\ldots})cc23)c1} \\
\rowcolor{green!12}
62 & QPU-only & active & \(-0.420\) & \(0.022\) & \href{https://pubchem.ncbi.nlm.nih.gov/compound/134812699}{\texttt{COc1cc(N2CCN(C)C}}\newline\href{https://pubchem.ncbi.nlm.nih.gov/compound/134812699}{\texttt{{\ldots}cccc2)s1}} \\
\rowcolor{red!12}
69 & QPU-only & inactive & \(-0.090\) & \(0.039\) & \texttt{OCc1ccc(-c2ccc3n}\newline\texttt{{\ldots})c3c2)o1} \\
\rowcolor{green!12}
87 & QPU-only & active & \(-0.239\) & \(0.324\) & \href{https://pubchem.ncbi.nlm.nih.gov/compound/71529733}{\texttt{CN(C)CCNC(=N)c1n}}\newline\href{https://pubchem.ncbi.nlm.nih.gov/compound/71529733}{\texttt{{\ldots}5)c3c2s1}} \\
17 & classical-only & active & \(0.121\) & \(-0.024\) & \texttt{COc1ccc2c(c1)CC(}\newline\texttt{{\ldots})ncnc1-2} \\
19 & classical-only & active & \(0.180\) & \(-0.661\) & \texttt{COc1ccc2c(c1)[nH}\newline\texttt{{\ldots}c(C)cc12} \\
50 & classical-only & active & \(0.015\) & \(-0.321\) & \texttt{COc1cc(-c2cc(=O)}\newline\texttt{{\ldots}o2)ccc1O} \\
61 & classical-only & inactive & \(0.631\) & \(-0.102\) & \texttt{CCn1c(-c2nonc2N)}\newline\texttt{{\ldots}NCC3)c21} \\
72 & classical-only & active & \(0.206\) & \(-0.003\) & \texttt{COc1ccc2c(c1)[nH}\newline\texttt{{\ldots}(CO)cc12} \\
74 & classical-only & inactive & \(0.270\) & \(-0.214\) & \texttt{CC[C@@H](CO)Nc1c}\newline\texttt{{\ldots}C)C)c2n1} \\
79 & classical-only & active & \(0.999\) & \(-0.068\) & \texttt{COc1ccc2c3ccnc(C}\newline\texttt{{\ldots})=O)c2c1} \\
80 & classical-only & inactive & \(0.322\) & \(-0.189\) & \texttt{COc1ccc2[nH]c3c(}\newline\texttt{{\ldots}C)c3c2c1} \\
81 & classical-only & active & \(0.704\) & \(-0.144\) & \texttt{NS(=O)(=O)c1ccc(}\newline\texttt{{\ldots}c3c2)cc1} \\
\bottomrule
\end{tabular*}
\end{table}

Recovering a patented analogue or an already-published probe, among the two labeled actives from the original split, says the disagreement touches known chemistry and which molecules deserve closer inspection.

Index 87. Thiazolo[5,4-\(f\)]quinazoline 8d~\cite{foucourt2014a}, the same structure as compound 33 of US~9{,}446{,}044~\cite{leblond2016}. Decision function on the table \(-0.24\); on the QPU \(+0.32\). The paper reports DYRK1A IC50 \(1.05\,\mu\mathrm{M}\); the patent's table reports 431~nM, under an assay protocol that does not necessarily match this curation's — this work's binary label does not depend on which of the two figures is correct, only on the \(\leq 1\,\mu\mathrm{M}\) cutoff. The series includes the much more potent leads EHT~1610 and EHT~5372~\cite{foucourt2014b,coutadeur2015}. What the QPU model recovers is a patented analogue, not the clinical-grade lead.

Index 62. XMD14-124, the ligand of PDB~6EIQ~\cite{czarna2018,rcsb6eiq}. Decision function on the table \(-0.42\); on the QPU \(+0.02\). Annotated affinity: \(K_{\mathrm{i}}\) 383~nM, IC50 800~nM. It is an academic probe, not a commercial asset.

The four QPU-only inactives are the cost of that extra list. They are not named as discoveries.

\subsection{kNN raises the ranking; the pair in question does not change}
\label{sec:results-knn}

On the same QPU vector, 213 expectation values, kNN raises the ROC-AUC from 0.8255 to 0.8583. The same effect, smaller, shows up on the table: kNN 0.8908 against RBF-SVM 0.8790. On the MPS simulator the RBF-SVM remains the best classifier, 0.8895.

The RBF-SVM-against-RBF-SVM comparison isolated the representation's effect. On this split the QPU alone does not beat the table; on two of the ten new partitions it does. Those partitions, and their extension to a hundred, are explored further below.

\subsection{Combining raises the ranking and does not rescue both actives at once}
\label{sec:results-combo}

Concatenating the 1080 columns with the 213 expectation values and classifying with kNN gives ROC-AUC 0.9022, F1 0.8679: \(+0.0115\) over the best single representation (table, 0.8908). Stacking on the kNN-classical and kNN-QPU scores reaches the same ROC-AUC, 0.9022, with F1 0.8598. The meta-classifier's coefficients are 2.5586 for the table's score and 1.9313 for the QPU's. The QPU vector weighs less; its coefficient is not zero.

The exact tie in ROC-AUC is plausible with 90 test molecules: AUC depends on relative order, not on the score's value. The gain shows up in both recipes; it comes from the joint signal.

The same concatenation with the MPS simulator's vector, instead of the QPU's, lands at 0.8865. The effect is stronger with the measured QPU vector.

No combined model recovers both QPU-only actives at once. Concatenation keeps XMD14-124 and loses 8d. Stacking keeps 8d and loses XMD14-124. Both retain the six actives the table alone proposed. Of the 28 molecules neither the table nor the QPU proposed separately, two are labeled active; neither combined model rescues them. Instead, both add the same two inactives to the list. Table~\ref{tab:combo-inactives} names them.

\begin{table}[H]
\centering
\caption{Inactives added by both combinations.}
\label{tab:combo-inactives}
\parbox{\columnwidth}{\footnotesize Concatenation and stacking add these two laboratory inactives. \(f\) is the signed decision function of the RBF-SVM for each representation. A negative value is that list's inactive side.\vspace{0.4em}}
\footnotesize
\setlength{\tabcolsep}{4pt}
\begin{tabular*}{\columnwidth}{@{\extracolsep{\fill}}rlcc p{0.42\columnwidth}@{}}
\toprule
Index & Label & \(f\) table & \(f\) QPU & SMILES \\
\midrule
8 & inactive & \(-0.268\) & \(-0.119\) & \url{COc1ccc(-c2cc3c(-c4ccccc4)c[nH]c3cn2)cc1} \\
76 & inactive & \(-0.036\) & \(-0.001\) & \url{Cc1c2c(cc[n+]1Cc1ccccc1)c1ccc(OCC3CCCCC3)cc1n2CC1CCCCC1} \\
\bottomrule
\end{tabular*}
\end{table}


\subsection{8d and XMD14-124 reappear when the split changes}
\label{sec:results-splits}

In the ten new partitions, 8d fell into test twice and in both the only proposer was the QPU's RBF-SVM (2/2). XMD14-124 fell into test once and was likewise proposed only by the QPU (1/1). With the available sample, the pattern repeats whenever the molecule appears in test.

The table's RBF-SVM beats the QPU's in ROC-AUC in 8 of the 10 partitions. The preprint's split is on that side. In 2 partitions, seeds 4 and 7, the QPU wins (0.8575 versus 0.8510; 0.8500 versus 0.8245). The classical ranking advantage is consistent but not absolute.

The QPU-only set ranges from 3 to 19 molecules (mean \(11.0 \pm 5.3\)); the classical-only set, from 3 to 8 (mean \(5.9 \pm 2.2\)). The preprint's split, 6 QPU-only, sits at the low end of that range.

Table~\ref{tab:split-robustness} lists the ten partitions: the ROC-AUC on each side, the size of each exclusive set, and whether 8d or XMD14-124 fall in that test.

\begin{table}[H]
\centering
\caption{Ten stratified 264/90 partitions.}
\label{tab:split-robustness}
\parbox{\columnwidth}{\small Each row is one repartition of the 354 compounds. The two ROC-AUC columns are that side's ranking on the row's 90. QPU and Classical are the size of the exclusive lists. The last two columns record whether 8d and XMD14-124 fall in that test and, when they do, which list proposes them.\vspace{0.55em}}
\footnotesize
\setlength{\tabcolsep}{3pt}
\begin{tabular*}{\columnwidth}{@{\extracolsep{\fill}}ccccccc@{}}
\toprule
Seed & \thh{ROC-AUC\\Table} & \thh{ROC-AUC\\QPU} & QPU & Classical & 8d & XMD14-124 \\
\midrule
0 & 0.8275 & 0.8065 & 17 & 3 & not in test & not in test \\
1 & 0.8605 & 0.8190 & 11 & 8 & QPU-only   & not in test \\
2 & 0.8060 & 0.7850 &  7 & 8 & not in test & not in test \\
3 & 0.8870 & 0.8175 & 14 & 3 & not in test & QPU-only \\
4 & 0.8510 & 0.8575 & 14 & 4 & not in test & not in test \\
5 & 0.8730 & 0.8480 & 13 & 8 & not in test & not in test \\
6 & 0.8855 & 0.8110 &  5 & 8 & not in test & not in test \\
7 & 0.8245 & 0.8500 &  7 & 7 & not in test & not in test \\
8 & 0.8475 & 0.8125 &  3 & 6 & not in test & not in test \\
9 & 0.8625 & 0.8425 & 19 & 4 & QPU-only   & not in test \\
\bottomrule
\end{tabular*}
\end{table}

A later extension, with the same protocol over a hundred new partitions instead of ten, replaces the raw count with a conditional rate and confidence interval: 8d falls into test 21 times and the QPU alone proposes it in 19 of those (90.5\,\%, 95\,\% Wilson CI [71.1\,\%, 97.3\,\%]); XMD14-124 falls 27 times and the QPU alone proposes it in 20 (74.1\,\%, CI [55.3\,\%, 86.8\,\%]). The QPU wins ROC-AUC in 21 of the 100 partitions (CI [14.2\,\%, 30.0\,\%]) — the same reading as the ten, now with a sample ten times larger: the classical ranking advantage is consistent but not absolute.

\subsection{Two more actives, recurrent in the new partitions}
\label{sec:results-new-candidates}

Cataloging which other active molecules appear as QPU-only across the hundred partitions found, besides 8d and XMD14-124, two molecules that recur more often than those two: QPU-only-active in 34 of the 34 times each falls into test (100\,\%, 95\,\% Wilson CI [89.8\,\%, 100\,\%]), against 90.5\,\% for 8d and 74.1\,\% for XMD14-124. Neither was in the original 90-molecule split; all their evidence comes from the new partitions.

\begin{figure}[H]
\centering
\includegraphics[width=0.72\columnwidth]{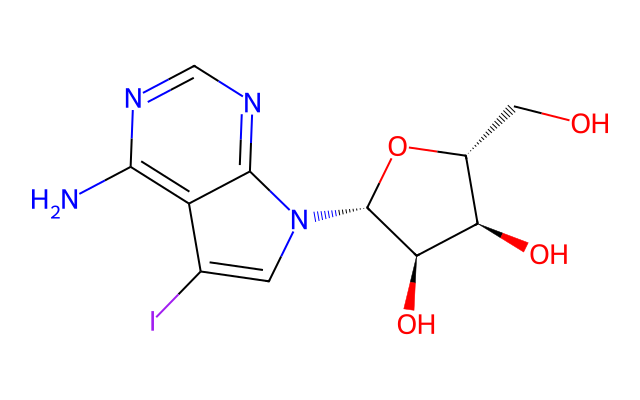}
\caption{5-Iodotubercidin, QPU-only in all 34/34 partitions where it falls into test.}
\label{fig:mol-5it}
\vspace{0.3em}
\parbox{\columnwidth}{\small A kinase inhibitor identified by kinome profiling against 253 human kinases; DYRK1A IC50 14~nM, more potent than harmine (80--300~nM). PubChem CID~97297~\cite{dirice2016}.\vspace{0.55em}}
\end{figure}

\begin{figure}[H]
\centering
\includegraphics[width=0.72\columnwidth]{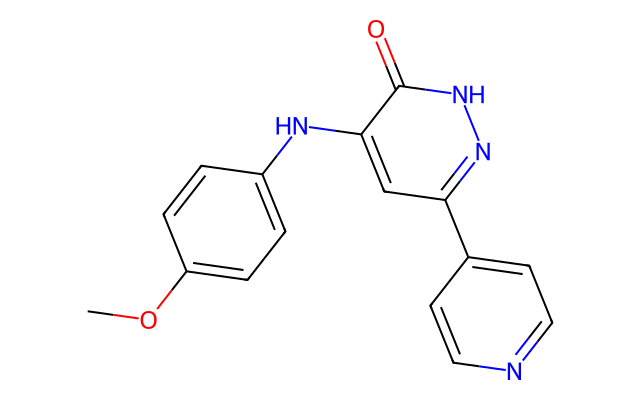}
\caption{Second candidate, QPU-only in all 34/34 partitions where it falls into test.}
\label{fig:mol-cand2}
\vspace{0.3em}
\parbox{\columnwidth}{\small DYRK1A IC50 610~nM, confirmed in the PubChem bioassay against UniProt target Q13627. PubChem CID~118715638~\cite{bruel2014,pubchem118715638}.\vspace{0.55em}}
\end{figure}

Both are in the training set of the canonical split (seed 42, \(y=1\)), not in the 90-molecule test set: neither was ever a candidate in the paper's main comparison. The classical applicability domain tells the two apart: 5-Iodotubercidin falls outside the table's domain (maximum Tanimoto 0.167 against train; the nucleoside has no close analogue) and inside the QPU vector's domain; the second candidate falls inside the domain in both representations (Tanimoto 0.854). The rest of the margin/\(\varepsilon\)-set pipeline was not run on these two with the same detail as for 8d/XMD14-124.

Cataloged together, the method flagged 46 distinct active molecules as QPU-only in at least one of the hundred partitions: 8d, XMD14-124, and 44 more. Of those 44, five recur twenty or more times, ten between ten and nineteen, and nineteen only one to three times, a range consistent with the sampling noise expected over a hundred repetitions. This work investigates four actives in depth in total: two from the original split, 8d and XMD14-124, and two more from this extension, the catalog's most recurrent ones (Figures~\ref{fig:mol-5it} and~\ref{fig:mol-cand2}). The remaining 42 candidates were neither identified nor investigated.

\subsection{8d and XMD14-124 interpolate}
\label{sec:results-ad}

At the 95th percentile, declared on train, 81 of the 90 test molecules interpolate on the table, 86 on the QPU vector, 82 on the concatenation, and 80 in the intersection of the two bases. The kNN-distance domain does not expel the test set. The Tanimoto \(\geq 0.70\) check is stricter: 52 of 90.

8d (index 87) and XMD14-124 (index 62) interpolate on all four descriptor flags, and in Tanimoto too (0.851 and 0.754). XMD14-124 sits close to the QPU space's threshold, distance 19.05 against 19.93, and does not cross it. At the 90th and 99th percentiles, 8d and XMD14-124 remain inside on the table, on the QPU vector, and on the concatenation.

The table had training neighbors similar to those two molecules. It still classified them as inactive. The disagreement is one of boundary within a region both cover.

In the concatenation, the subset inside the domain (\(n=82\)) has recall 0.938 and MCC 0.699; the subset outside (\(n=8\)) has recall 0.50 and MCC 0.333. The domain discriminates, on a small outside sample. On the QPU the 4 outside are all inactive: the Cooper statistic is undefined outside. On the table the 9 outside do not collapse (\textit{accuracy} 0.889); \(n=9\) does not license saying the domain cleans up the error.

It covers Principle 3 on this split, with the method declared.

\subsection{Where each model draws the line}
\label{sec:results-boundary}

The table's RBF-SVM leaves out 8d and XMD14-124. The QPU's lets them in. The pair again flags the same six QPU-only (indices 24, 54, 59, 62, 69, 87) and the same nine classical-only.

The \(\varepsilon\)-set are the classifiers whose train F1 sits at most 0.03 below that space's winner. With \(\varepsilon = 0.03\), all five families qualify on both sides. On the table, 0 of 5 propose 8d and 0 of 5 propose XMD14-124. On the QPU, 5 of 5 propose 8d and 4 of 5 propose XMD14-124; the QPU's logistic regression leaves XMD14-124 out. Same family on both sides: 8d is QPU-only in all five; XMD14-124 in four of five.

Lowering \(\varepsilon\) does not change that reading. At 0.02, three table classifiers remain (0/3 and 0/3) and two QPU ones (2/2 and 2/2). At 0.01, the same three table classifiers remain (0/3 and 0/3) and only the random forest on the QPU (1/1 and 1/1).

XMD14-124 on the QPU has margin 0.022, 2.2nd percentile of 90. 8d on the table: 0.239, 20.2nd percentile. XMD14-124 on the table: 0.420, 31.5th percentile. 8d on the QPU: 0.324, 46.1st percentile. In a region both cover, one model places them on one side and the other on the opposite side.

For a visual check, we use these diagrams and place 8d and XMD14-124 on each representation's decision surface. The 2-D PCA explains 23.8\,\% of the variance on the table and 11.4\,\% on the QPU.

\begin{figure}[H]
\centering
\includegraphics[width=\columnwidth]{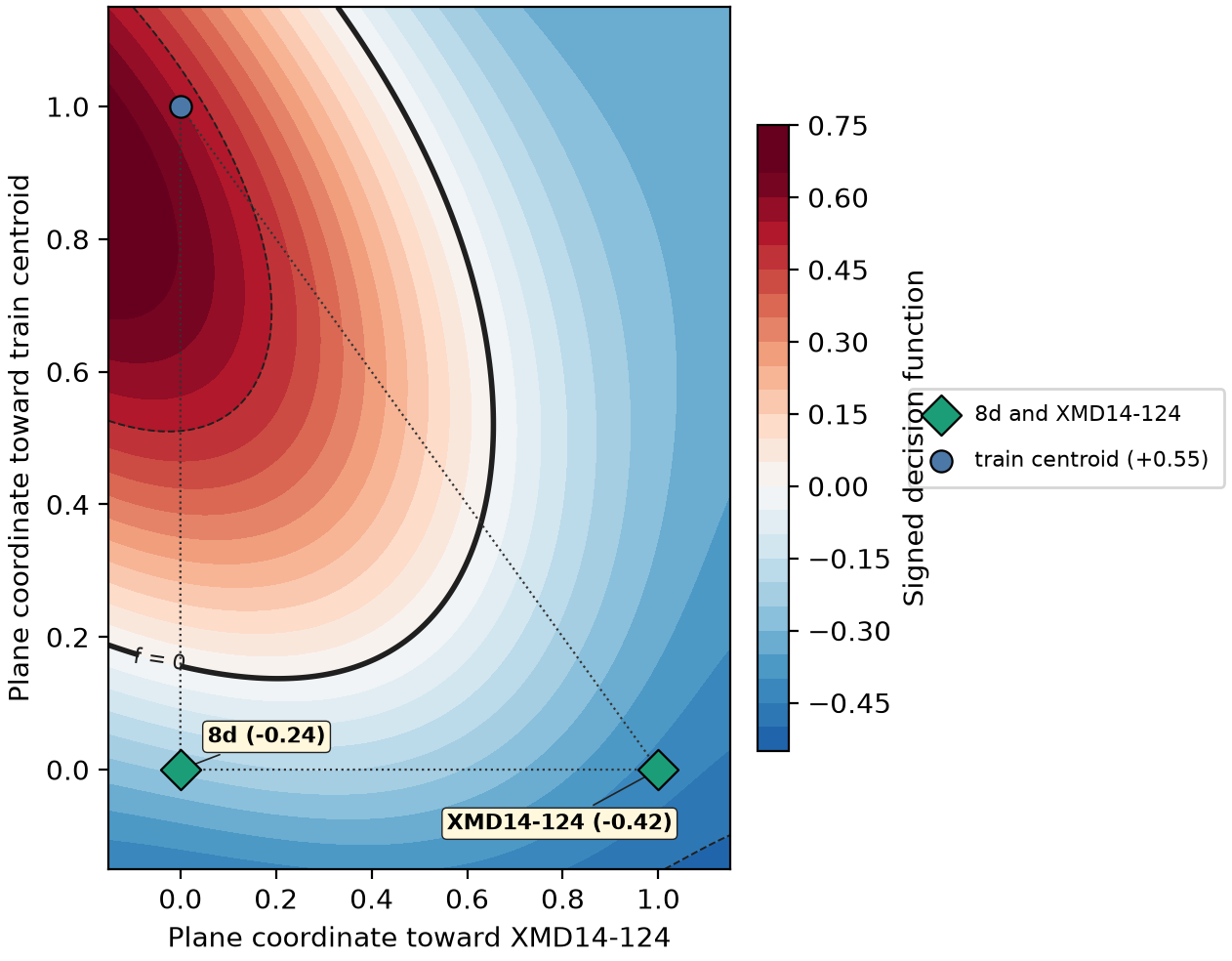}
\caption{Somepalli cut of the 1080-dimensional RBF-SVM.}
\label{fig:somepalli-classical}
\parbox{\columnwidth}{\small The plane passes through 8d, XMD14-124, and the train centroid. Color is the signed decision function. The solid line is \(f=0\). Both molecules fall on the inactive side~\cite{somepalli2022}.\vspace{0.4em}}
\end{figure}

Figure~\ref{fig:somepalli-classical} is the 1080-dimensional RBF-SVM in the plane through 8d, XMD14-124, and the train centroid: both fall on the inactive side.

\begin{figure}[H]
\centering
\includegraphics[width=\columnwidth]{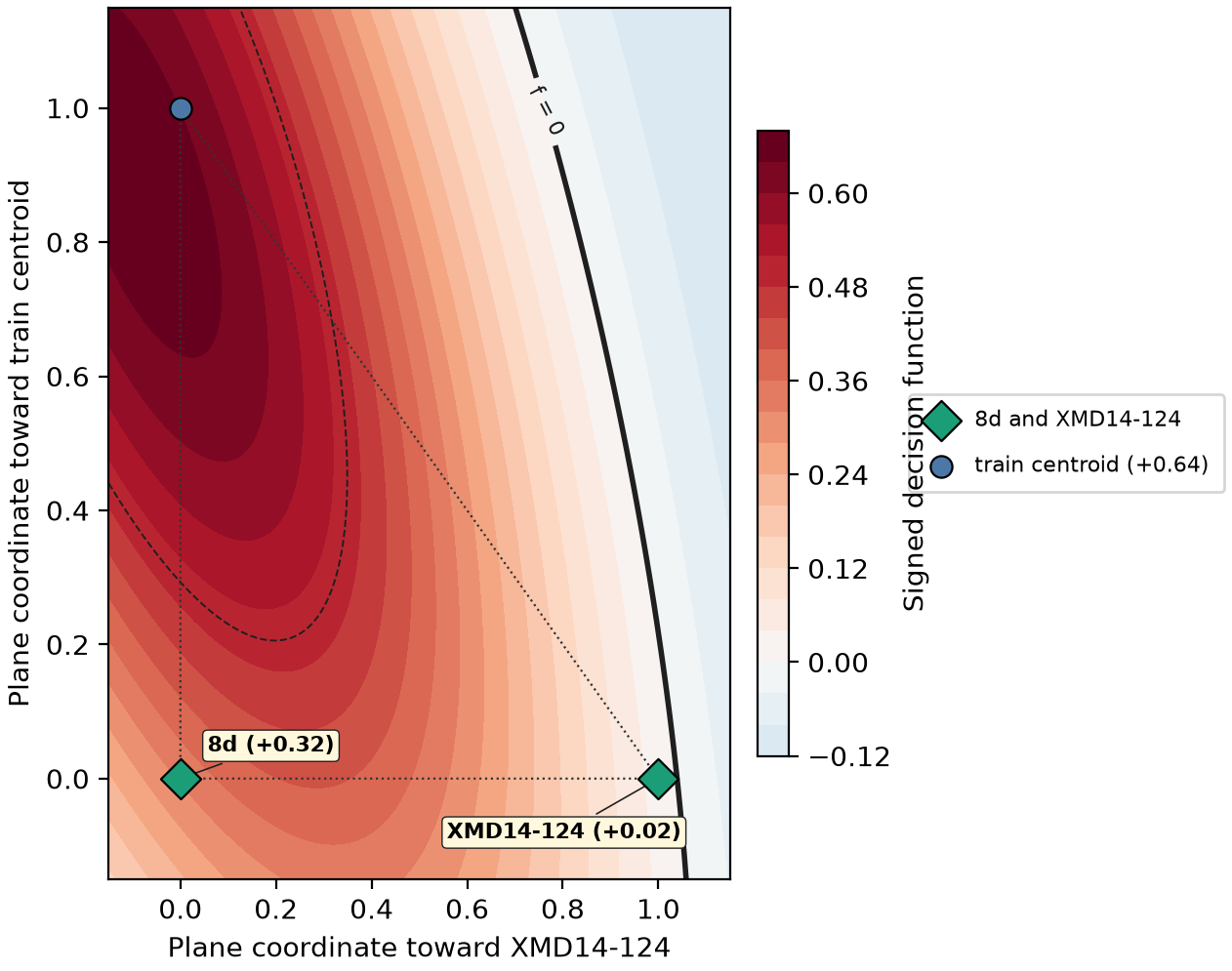}
\caption{The same plane, 213-dimensional RBF-SVM.}
\label{fig:somepalli-qpu}
\parbox{\columnwidth}{\small 8d is active (\(f=+0.32\)). XMD14-124 sits on the line (\(f=+0.02\)).\vspace{0.4em}}
\end{figure}

Figure~\ref{fig:somepalli-qpu} is the same plane for the 213-dimensional RBF-SVM: 8d turns active and XMD14-124 sits on the line.

\begin{figure}[H]
\centering
\includegraphics[width=\columnwidth]{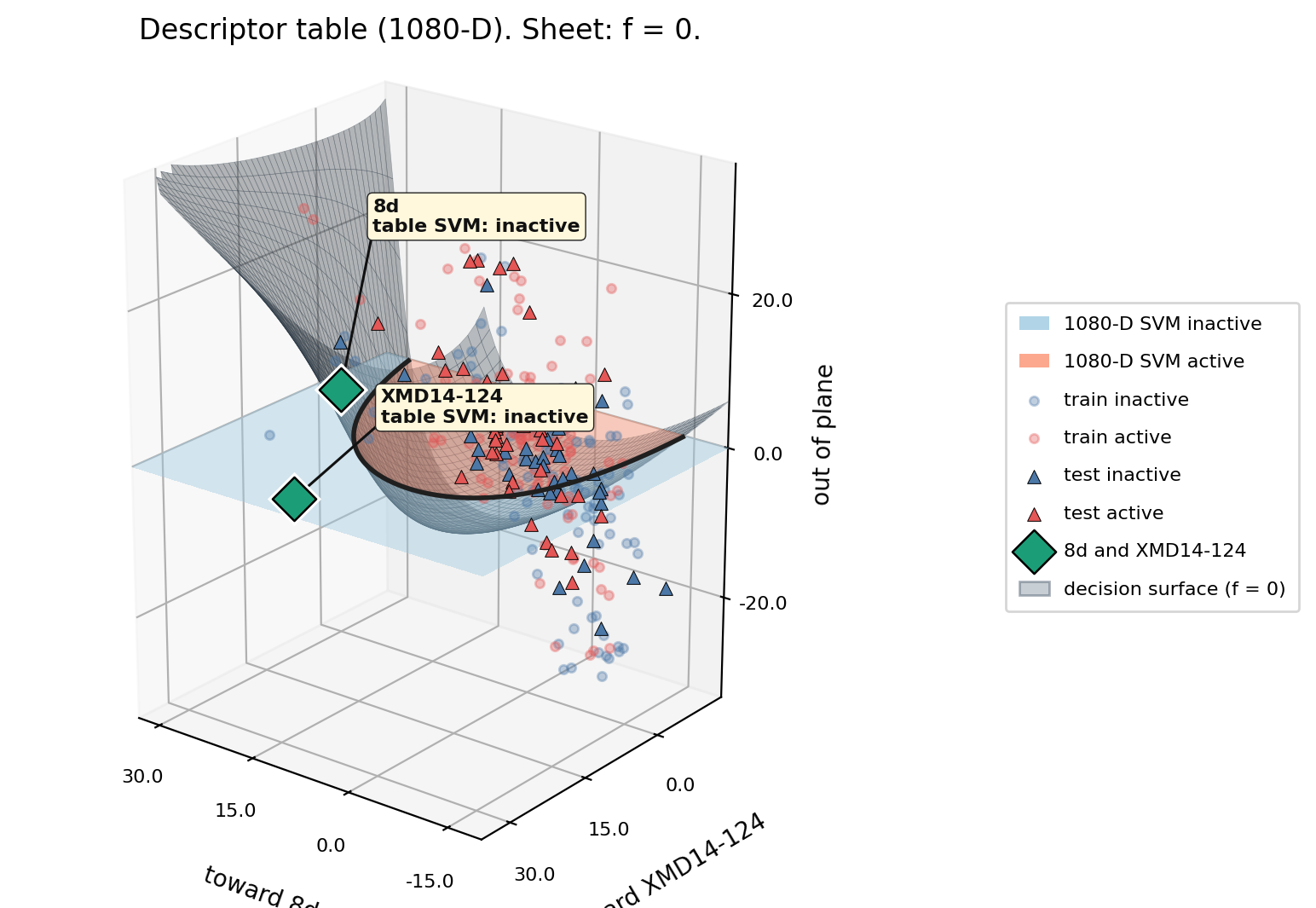}
\caption{1080-dimensional RBF-SVM, cut in three dimensions.}
\label{fig:pca3-classical}
\parbox{\columnwidth}{\small The sheet is \(f=0\) in the cut through 8d, XMD14-124, and the train centroid. Both molecules fall on the inactive side. It is the pair's surface, not an SVM refit on three components.\vspace{0.4em}}
\end{figure}

Figure~\ref{fig:pca3-classical} is that same sheet in three dimensions. It is the pair's surface.

\begin{figure}[H]
\centering
\includegraphics[width=\columnwidth]{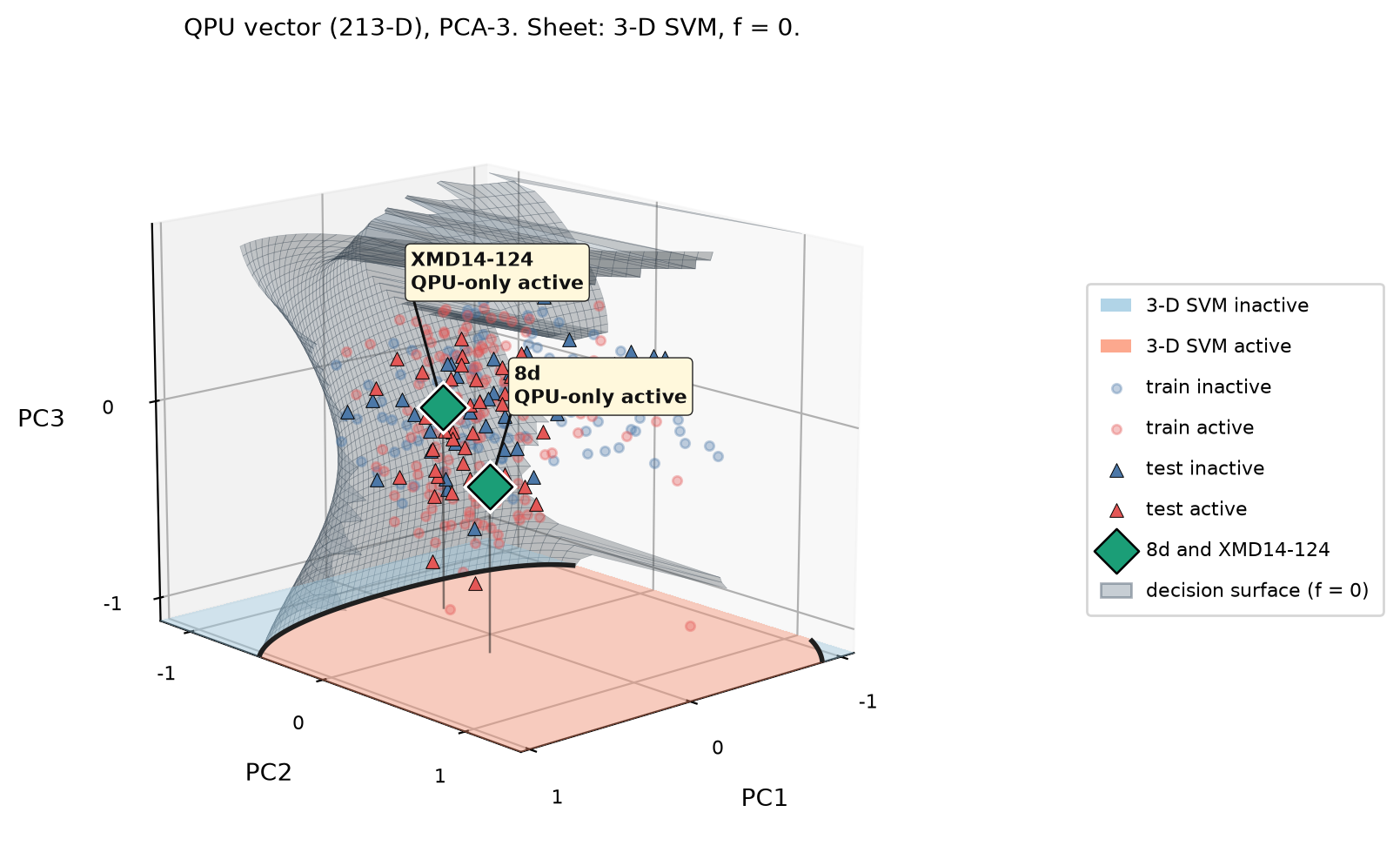}
\caption{QPU vector, PCA-3.}
\label{fig:pca3-qpu}
\parbox{\columnwidth}{\small The sheet is \(f=0\) of an RBF-SVM fit in this reduced space. The green diamonds are the QPU-only actives. It is not the 213-dimensional surface.\vspace{0.4em}}
\end{figure}

Figure~\ref{fig:pca3-qpu} is the QPU vector in a PCA-3 SVM. This drawing is not the pair's surface.

\begin{figure}[H]
\centering
\includegraphics[width=\columnwidth]{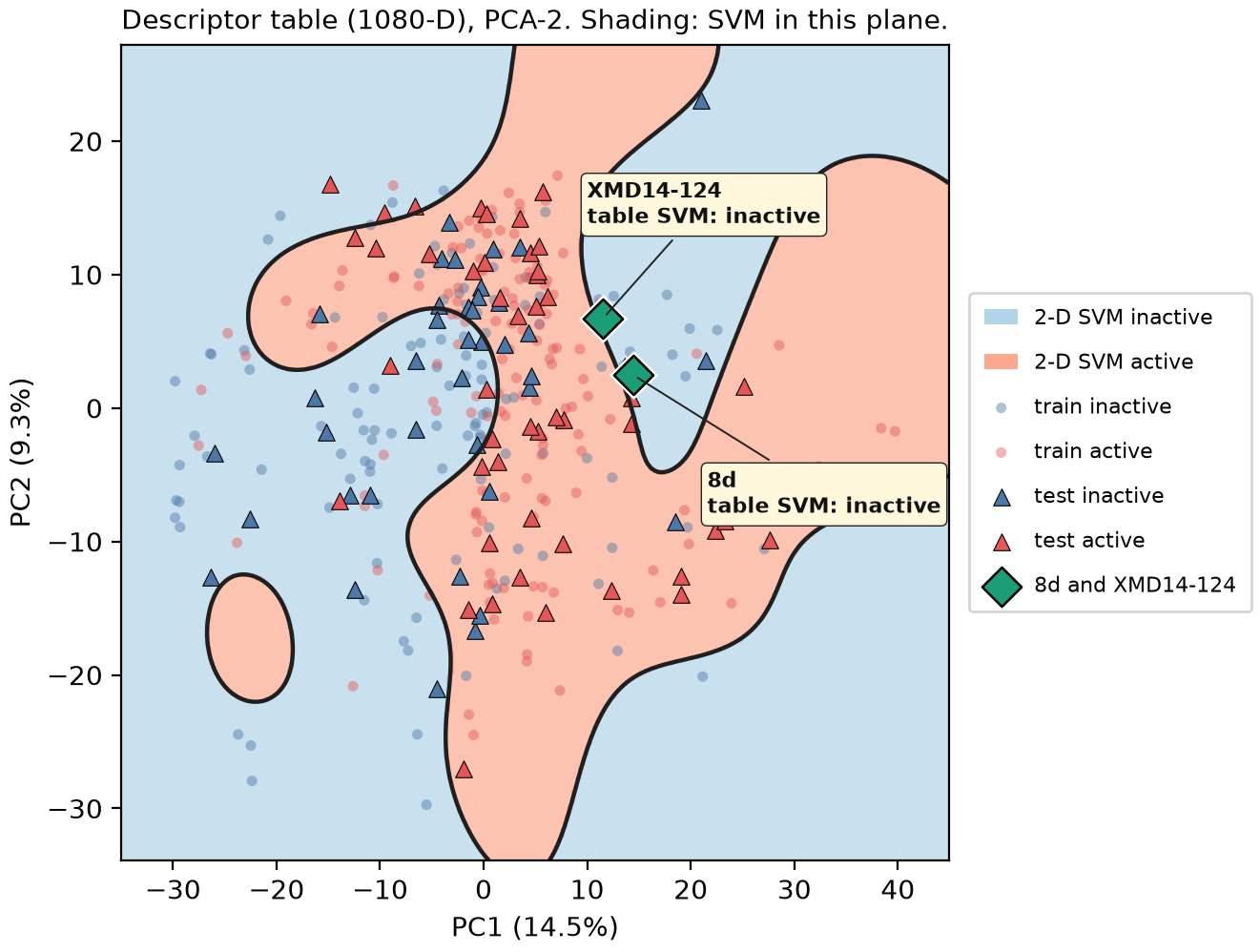}
\caption{Descriptor table, PCA-2.}
\label{fig:pca2-classical}
\parbox{\columnwidth}{\small Blue is the SVM's inactive side in two dimensions. Salmon is the active side. The green diamonds are 8d and XMD14-124, both inactive under the 1080-dimensional SVM. This drawing is not that surface.\vspace{0.4em}}
\end{figure}

Figure~\ref{fig:pca2-classical} is the two-dimensional illustration of the descriptor table. This drawing is not the 1080-dimensional surface.

\begin{figure}[H]
\centering
\includegraphics[width=\columnwidth]{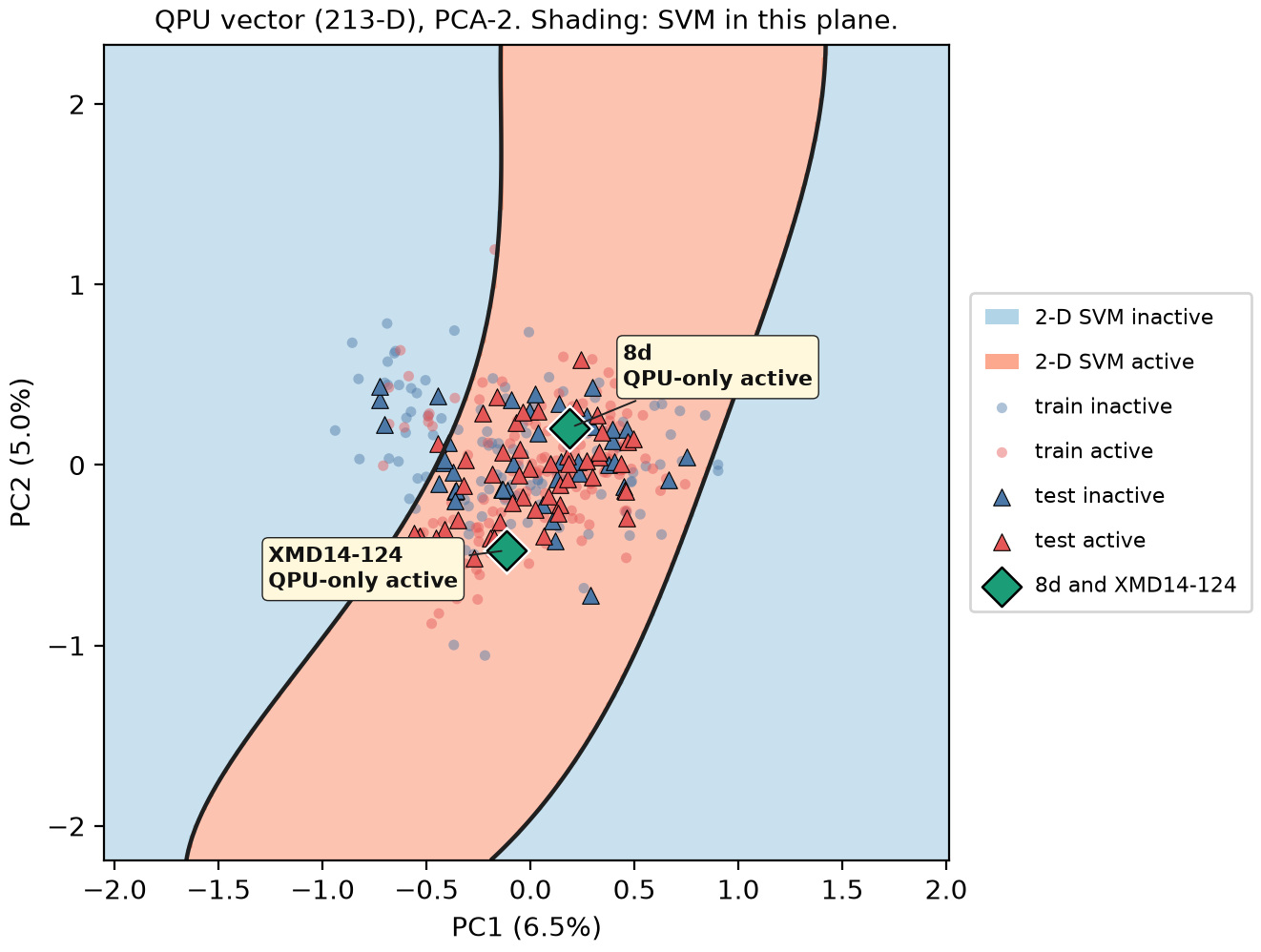}
\caption{QPU vector, PCA-2.}
\label{fig:pca2-qpu}
\parbox{\columnwidth}{\small Same color key as Fig.~\ref{fig:pca2-classical}. 8d and XMD14-124 fall in the active band. This drawing is not the 213-dimensional surface.\vspace{0.4em}}
\end{figure}

Figure~\ref{fig:pca2-qpu} is the two-dimensional illustration of the QPU vector. This drawing is not the 213-dimensional surface.

\subsection{Each block's contribution is comparable and visible molecule by molecule}
\label{sec:results-shapley}

Four readings answer different questions about the same two representations and are consistent with each other, not convergent in a formal statistical sense: Shapley splits the aggregate contribution to the ranking, Pearson/Spearman/\(Q\) Yule's quantify agreement between models, SHAP attributes the vote per molecule, and mutual information measures how much each component contributes on its own. Across all four, the table's block and the QPU's contribute comparable magnitudes, with the table's somewhat larger, and the two signals are correlated without being redundant.

The Shapley value splits the joint ROC-AUC between the two blocks. \(v(\emptyset)=0.5\) is chance; \(v(\mathrm{table})=0.8908\), \(v(\mathrm{QPU})=0.8583\), \(v(\mathrm{both})=0.9022\). The split assigns 54.0\,\% to the table and 46.0\,\% to the QPU vector. The efficiency axiom checks out: \(0.2174 + 0.1849 = 0.4022\).

The two kNN's scores have Pearson \(r=0.7333\) and Spearman \(r=0.7224\). Yule's \(Q\) is 0.879~\cite{kuncheva2003}. The disagreement rate is 17.8\,\% (16 of 90). Both fail on the same 12 molecules (13.3\,\%). Those 12 fall outside the reach of any combination of these two bases.

SHAP on the meta-classifier attributes the combined vote, molecule by molecule. For 8d, the QPU's push (\(+0.313\)) flips the table's vote (\(-0.542\)) to active, a correct prediction. For XMD14-124, the same table vote (\(-0.542\)) gets \(+0.137\) from the QPU, score 0.727 against 8d's 0.818, and does not flip: an incorrect prediction, inactive. The table dominates the local attribution on 72 of 90 molecules; the QPU, on 18.

On the table, 290 of 1080 descriptors (26.9\,\%) have zero mutual information with the label. On the QPU vector, 2 of 71 qubits (2.8\,\%). Nearly every qubit contributes something, weak and spread out.

\subsection{Hardware noise does not explain all of the disagreement}

For the six QPU-only molecules, the difference between the MPS simulator's clean vector and the QPU vector on the 264 training molecules has mean absolute value 0.052 and standard deviation 0.071 per component; 49 of the 213 components show a systematic bias, and the correlation between neighboring components is weak on average (0.050), aside from a few isolated pairs (maximum 0.321).

Injecting that noise onto each of the six's clean vector, the crossing rate to the active side averages 38.6\,\% with the resampling model and 39.6\,\% with the independent Gaussian one; on the five contrast molecules, far from the boundary, the rate drops to 12.8\,\% and 12.2\,\%. Noise alone does not reproduce the crossing rate observed in the six.

Two distinct mechanisms separate the six. For indices 24 and 59, the score is already positive on the MPS simulator's clean vector (0.172 and 0.193): noise does not cross them, they were already on the active side. For 8d and XMD14-124, the clean score is near zero or negative (0.047 and \(-0.291\)) and the real jump measured on hardware, \(+0.277\) and \(+0.313\), is what crosses them: there, physical noise contributes signal beyond the clean representation.

\subsection{The ranking advantage does not reach conventional significance}

A paired \textit{bootstrap} of 2000 resamples over the 90 test molecules gives a 95\,\% confidence interval of [0.802, 0.941] for the table's ROC-AUC and [0.730, 0.908] for the QPU's; the interval for their difference is \([-0.012, 0.122]\) and includes zero. In 5.4\,\% of the resamples the QPU beats the table.

The DeLong test on the two ROC curves gives \(z=1.568\), \(p=0.117\): the hypothesis that both representations have the same ranking capacity is not rejected.

On the active/inactive composition of QPU-only (2/6) against classical-only (6/9), Fisher's exact test gives an odds ratio of 0.25 with \(p=0.315\); the Wilson intervals for each rate, [9.7\,\%, 70.0\,\%] for QPU-only and [35.4\,\%, 87.9\,\%] for classical-only, overlap heavily. Neither reading is significant at this sample size.

\section{Discussion}
\label{sec:discussion}

The two RBF-SVMs' lists differ. 8d and XMD14-124 fall within the training neighborhood of both representations; when they land in test, in this sample, the measured-vector model marks them active and the table leaves them out. On the table, classifiers with F1 close to the winner do not mark them either. On the measured vector, 8d is active across all five families, and XMD14-124 is active in four, right at the threshold, with that vector's logistic regression leaving it out. An active that does not move to experimental evaluation leaves its chemical series out of the next assay, and a higher area under the curve on the table would have left both of them out.

We cannot reject the hypothesis that both models have equivalent ranking capacity (DeLong p=0.117), which is consistent with the observed disagreement not coming from one model being globally superior, but from where each representation draws its decision boundary.

Two distinct mechanisms explain the six QPU-only molecules. For indices 24 and 59, the noise-free simulation of the same EME circuit already places them on the active side before any hardware noise enters: it is a property of the representation itself. For XMD14-124 and 8d, the vector measured on hardware moves the score more than the noise calibrated from this backend's real data typically produces: there, physical noise contributes signal beyond the clean representation.

The union of the two lists pays four extra inactives and leaves out the six actives the QPU does not see. It is a plate policy.

The controls, same vectors, no new submissions, close off rival readings. kNN raises the QPU vector's ranking; the RBF-SVM pair was not that vector's ceiling. On the preprint's split the QPU alone does not beat the table; on 21 of a hundred new partitions it does.

Concatenating or stacking reaches the same ranking ceiling. No combination recovers 8d and XMD14-124 at once; both retain the actives the table alone saw. The QPU's score carries weight in the stacking: the vector contributes. Concatenation keeps XMD14-124 and loses 8d; stacking the reverse.

Over a hundred new partitions, 8d falls into test 21 times and in 19 of those only the QPU's RBF-SVM proposes it (90.5\,\%, 95\,\% Wilson CI [71.1\,\%, 97.3\,\%]); XMD14-124 falls 27 times and in 20 only the QPU proposes it (74.1\,\%, CI [55.3\,\%, 86.8\,\%]). The lists disagree again across those partitions. SHAP attributes the combined vote: the QPU's push shows up more in 8d. Both surfaces of the original pair remain without a mechanism. Twelve of the 90 are missed by both bases: a ceiling on combining these two models.

Several of the molecules that recur as QPU-only-active across those hundred partitions share 8d's thiazoloquinazoline core, consistent with the disagreement concentrating in a chemical series, a pattern flagged here but not pursued in this work (see supplementary material).

Two of those recurrent candidates (Figures~\ref{fig:mol-5it} and~\ref{fig:mol-cand2}, 5-Iodotubercidin and the second candidate) turn out to be already-published DYRK1A inhibitors that the classical table never proposed in any partition: the QPU discovers where the classical representation ignores. This work is not a virtual screening exercise and was not designed to validate itself as one, and the rest of that same catalog's candidates fall outside its scope: cataloging and investigating them in depth is future work, not part of this paper.

Outside the scope of this study: clinical benefit, a general quantum advantage in drug discovery, a dossier-grade QSAR, and the QMKL AUC figures from the earlier draft. DYRK1A is the reason the assay exists~\cite{coutadeur2015}. Neither of the two recovered structures is an approved drug.

\section{Conclusion}
\label{sec:conclusion}

Over the same 90 DYRK1A molecules, the two RBF-SVMs propose different lists: six compounds only on the measured vector, among them 8d and XMD14-124, and six actives only on the table.

Those two actives do not share the margin on the measured vector. 8d is active across all five families; XMD14-124 sits right at the threshold. The area under the curve, 0.8255 against 0.8790, orders the test, without conventional statistical significance at this sample size (paired \textit{bootstrap} and DeLong, p=0.117), and experimental evaluation depends on the label each model assigns.

Concatenation and stacking reorder that ranking. In this partition each combination retains one of those two actives. The mechanism behind the two surfaces falls outside these experiments.

The same disagreement reappears across a hundred new partitions, where two additional recurrent candidates also emerge, and hardware noise calibrated from real data does not by itself reproduce the crossing rate observed in at least two of the six QPU-only molecules.

The answer to this work's question is yes: over the same DYRK1A test set, the classical descriptor table and the vector measured on the quantum processor propose different candidate lists for experimental evaluation, and that discrepancy survives a change of split and is not fully explained as hardware noise. Which of the two representations deserves more trust for deciding what moves to assay remains unresolved: neither dominates the other at conventional statistical significance, and the boundary mechanism that separates them remains open for future work.

\section*{Acknowledgment}

We thank the team at \href{https://www.globant.com/}{$\mathtt{Globant}$} for access to their computing capabilities, the support of their infrastructure, and their patience throughout this research.

Special thanks to \href{https://ibm.com/}{$\mathtt{IBM}$} for their guidance, suggestions, and patience in answering our questions.

We also thank \href{https://haiqu.ai/}{$\mathtt{Haiqu}$} and their technical team for their advice and support~\cite{haiqueme}. The extra weekend motivation helped close this work in record time.

Our thanks to \href{http://protoqsar.com/}{$\mathtt{ProtoQSAR}$} and their research team for providing the dataset~\cite{zdrazil2024}, which served as a reference and made it possible to continue this study.

We thank the nodes of the \href{https://ripaisc.net/}{$\mathtt{RIPAISC}$} network for the intellectual exchange. Their comments, by questioning and refining the methodology, strengthened this research's results.

This research was carried out within the framework of the Cervera Network of Excellence for research in quantum technologies for an experimental infrastructure oriented toward technological sovereignty in strategic settings ($\mathtt{CER-20251020\,|\,QUIXOTE}$). It was funded by the Ministry of Science, Innovation and Universities through $\mathtt{CDTI}$, and by the Ibero-American Program for Science and Technology for Development (CYTED) through RIPAISC, the Ibero-American Network for the Advancement of Quantum Software Engineering (525RT0174).

Finally, we thank the \href{https://qnow.tech/}{$\mathtt{QNOW}$} technical team for their contribution to the scaffolding, execution, and research of this project.

\bibliographystyle{IEEEtran}
\bibliography{references}

@article{cherkasov2014,
  author  = {Artem Cherkasov and Eugene N. Muratov and Denis Fourches and Alexandre Varnek and Igor I. Baskin and Mark Cronin and John Dearden and Paola Gramatica and Yvonne C. Martin and Roberto Todeschini and Viviana Consonni and Victor E. Kuz'min and Richard Cramer and Romualdo Benigni and Chihae Yang and James Rathman and Lothar Terfloth and Johann Gasteiger and Ann Richard and Alexander Tropsha},
  title   = {{QSAR} Modeling: Where Have You Been? Where Are You Going To?},
  journal = {Journal of Medicinal Chemistry},
  volume  = {57},
  number  = {12},
  pages   = {4977--5010},
  year    = {2014},
  doi     = {10.1021/jm4004285}
}

@article{coutadeur2015,
  author  = {S{\'e}verine Coutadeur and H{\'e}l{\`e}ne Benyamine and Laurence Delalonde and Catherine de Oliveira and Bertrand Leblond and Alicia Foucourt and Thierry Besson and Anne-Sophie Casagrande and Thierry Taverne and Ang{\'e}lique Girard and Matthew P. Pando and Laurent D{\'e}sir{\'e}},
  title   = {A novel {DYRK1A} (Dual specificity tyrosine phosphorylation-regulated kinase 1A) inhibitor for the treatment of {Alzheimer's} disease: effect on {Tau} and amyloid pathologies in vitro},
  journal = {Journal of Neurochemistry},
  volume  = {133},
  number  = {3},
  pages   = {440--451},
  year    = {2015},
  doi     = {10.1111/jnc.13018}
}

@article{foucourt2014a,
  author  = {Alicia Foucourt and Damien H{\'e}dou and Carole Dubouilh-Benard and Laurent D{\'e}sir{\'e} and Anne-Sophie Casagrande and Bertrand Leblond and Nad{\`e}ge Lo{\"a}ec and Laurent Meijer and Thierry Besson},
  title   = {Design and Synthesis of Thiazolo[5,4-f]quinazolines as {DYRK1A} Inhibitors, Part {I}},
  journal = {Molecules},
  volume  = {19},
  number  = {10},
  pages   = {15546--15571},
  year    = {2014},
  doi     = {10.3390/molecules191015546}
}

@article{foucourt2014b,
  author  = {Alicia Foucourt and Damien H{\'e}dou and Carole Dubouilh-Benard and Ang{\'e}lique Girard and Thierry Taverne and Anne-Sophie Casagrande and Laurent D{\'e}sir{\'e} and Bertrand Leblond and Thierry Besson},
  title   = {Design and Synthesis of Thiazolo[5,4-f]quinazolines as {DYRK1A} Inhibitors, Part {II}},
  journal = {Molecules},
  volume  = {19},
  number  = {10},
  pages   = {15411--15439},
  year    = {2014},
  doi     = {10.3390/molecules191015411}
}

@article{czarna2018,
  author  = {Anna Czarna and Jinhua Wang and Diana Zelencova and Yao Liu and Xianming Deng and Hwan Geun Choi and Tinghu Zhang and Wenjun Zhou and Jae Won Chang and Hanne Kildalsen and Ole Morten Seternes and Nathanael S. Gray and Richard A. Engh and Ulli Rothweiler},
  title   = {Novel Scaffolds for Dual Specificity Tyrosine-Phosphorylation-Regulated Kinase ({DYRK1A}) Inhibitors},
  journal = {Journal of Medicinal Chemistry},
  volume  = {61},
  number  = {17},
  pages   = {7560--7572},
  year    = {2018},
  doi     = {10.1021/acs.jmedchem.7b01847}
}

@article{havlicek2019,
  author  = {Vojt{\v{e}}ch Havl{\'i}{\v{c}}ek and Antonio D. C{\'o}rcoles and Kristan Temme and Aram W. Harrow and Abhinav Kandala and Jerry M. Chow and Jay M. Gambetta},
  title   = {Supervised learning with quantum-enhanced feature spaces},
  journal = {Nature},
  volume  = {567},
  number  = {7747},
  pages   = {209--212},
  year    = {2019},
  doi     = {10.1038/s41586-019-0980-2}
}

@patent{leblond2016,
  author = {Bertrand Leblond and Anne-Sophie Casagrande and Laurent D{\'e}sir{\'e} and Alicia Foucourt and Thierry Besson},
  title  = {{DYRK1} inhibitors and uses thereof},
  number = {9446044},
  nationality = {US},
  year   = {2016},
  note   = {Issued 20 Sep. 2016. Compound 33 matches {8d} of Foucourt et al.}
}

@misc{rcsb6eiq,
  author = {Anna Czarna and Jinhua Wang and Diana Zelencova and Yao Liu and Xianming Deng and Hwan Geun Choi and Tinghu Zhang and Wenjun Zhou and Jae Won Chang and Hanne Kildalsen and Ole Morten Seternes and Nathanael S. Gray and Richard A. Engh and Ulli Rothweiler},
  title  = {{DYRK1A} in complex with {XMD14-124}},
  howpublished = {RCSB {PDB} 6EIQ},
  year   = {2018},
  note   = {Ligand {B6Z}, 2.3~\AA. \url{https://www.rcsb.org/structure/6EIQ}}
}

@article{dirice2016,
  author  = {E. Dirice and D. Walpita and A. Vetere and B. C. Meier and S. Kahraman and J. Hu and V. Dan{\v{c}}{\'\i}k and S. M. Burns and T. J. Gilbert and D. E. Olson and P. A. Clemons and R. N. Kulkarni and B. K. Wagner},
  title   = {Inhibition of {DYRK1A} Stimulates Human \ensuremath{\beta}-Cell Proliferation},
  journal = {Diabetes},
  volume  = {65},
  number  = {6},
  pages   = {1660--1671},
  year    = {2016},
  doi     = {10.2337/db15-1127}
}

@article{bruel2014,
  author  = {A. Bruel and R. B{\'e}n{\'e}teau and M. Chabanne and O. Lozach and R. Le Guevel and M. Ravache and H. B{\'e}n{\'e}detti and L. Meijer and C. Log{\'e} and J.-M. Robert},
  title   = {Synthesis of New Pyridazino[4,5-b]indol-4-ones and Pyridazin-3(2{H})-one Analogs as {DYRK1A} Inhibitors},
  journal = {Bioorganic \& Medicinal Chemistry Letters},
  volume  = {24},
  number  = {21},
  pages   = {5037--5040},
  year    = {2014},
  doi     = {10.1016/j.bmcl.2014.09.017}
}

@misc{pubchem118715638,
  author       = {{PubChem}},
  title        = {{PubChem BioAssay AID 1165914}: Inhibition of {DYRK1A} (unknown origin), compound {CID 118715638}},
  howpublished = {National Center for Biotechnology Information},
  year         = {2026},
  note         = {Active, {IC50} 0.61~\(\mu\)M against {DYRK1A} (UniProt {Q13627}). \url{https://pubchem.ncbi.nlm.nih.gov/compound/118715638}}
}

@article{zdrazil2024,
  author  = {Barbara Zdrazil and Eloy Felix and Fiona Hunter and Emma J. Manners and James Blackshaw and Sybilla Corbett and Marleen de Veij and Harris Ioannidis and David Mendez Lopez and Juan F. Mosquera and Maria Paula Magari{\~n}os and Nicolas Bosc and Ricardo Arcila and Tevfik Kizil{\"o}ren and Anna Gaulton and A. Patr{\'i}cia Bento and Melissa F. Adasme and Peter Monecke and Gregory A. Landrum and Andrew R. Leach},
  title   = {The {ChEMBL} Database in 2023: a drug discovery platform spanning multiple bioactivity data types and time periods},
  journal = {Nucleic Acids Research},
  volume  = {52},
  number  = {D1},
  pages   = {D1180--D1192},
  year    = {2024},
  doi     = {10.1093/nar/gkad1004}
}

@article{neves2018,
  author  = {Bruno J. Neves and Rodolpho C. Braga and Cleber C. Melo-Filho and Jos{\'e} Te{\'o}filo Moreira-Filho and Eugene N. Muratov and Carolina Horta Andrade},
  title   = {{QSAR}-Based Virtual Screening: Advances and Applications in Drug Discovery},
  journal = {Frontiers in Pharmacology},
  volume  = {9},
  pages   = {1275},
  year    = {2018},
  doi     = {10.3389/fphar.2018.01275}
}

@misc{giraldo2025qqsar,
  author        = {Alejandro Giraldo and Daniel Ruiz and Mariano Caruso and Guido Bellomo},
  title         = {Quantum {QSAR} for drug discovery},
  year          = {2025},
  eprint        = {2505.04648},
  archivePrefix = {arXiv},
  primaryClass  = {quant-ph}
}

@inproceedings{giraldo2025q2sar,
  author        = {Alejandro Giraldo and Daniel Ruiz and Mariano Caruso and Javier Mancilla and Guido Bellomo},
  title         = {{Q2SAR}: A Quantum Multiple Kernel Learning Approach for Drug Discovery},
  booktitle     = {2025 IEEE CHILEAN Conference on Electrical, Electronics Engineering, Information and Communication Technologies (CHILECON)},
  pages         = {1--4},
  year          = {2025},
  address       = {Valpara{\'i}so, Chile},
  doi           = {10.1109/chilecon66915.2025.11476277},
  eprint        = {2506.14920},
  archivePrefix = {arXiv},
  primaryClass  = {quant-ph}
}

@article{renard2024,
  author  = {Xavier Renard and Thibault Laugel and Marcin Detyniecki},
  title   = {Understanding prediction discrepancies in classification},
  journal = {Machine Learning},
  volume  = {113},
  number  = {10},
  pages   = {7997--8026},
  year    = {2024},
  doi     = {10.1007/s10994-024-06557-4}
}

@misc{haiqueme,
  author       = {{Haiqu}},
  title        = {Feature embedding: {Entangled Manifold Embedding}},
  howpublished = {\url{https://docs.haiqu.ai/core_features/data_loading_isom}},
  year         = {2026},
  note         = {Accessed 17 Sep. 2026}
}

@techreport{oecd2007,
  author      = {{OECD}},
  title       = {Guidance Document on the Validation of {(Quantitative)} Structure-Activity Relationship [{(Q)SAR}] Models},
  institution = {OECD},
  series      = {OECD Series on Testing and Assessment},
  number      = {No. 69, ENV/JM/MONO(2007)2},
  year        = {2007},
  doi         = {10.1787/9789264085442-en},
  note        = {Crossref records the online edition as 2014-09-03}
}

@article{eriksson2003,
  author  = {Lennart Eriksson and Joanna Jaworska and Andrew P. Worth and Mark T. D. Cronin and Robert M. McDowell and Paola Gramatica},
  title   = {Methods for reliability and uncertainty assessment and for applicability evaluations of classification- and regression-based {QSARs}},
  journal = {Environmental Health Perspectives},
  volume  = {111},
  number  = {10},
  pages   = {1361--1375},
  year    = {2003},
  doi     = {10.1289/ehp.5758}
}

@article{chicco2020,
  author  = {Davide Chicco and Giuseppe Jurman},
  title   = {The advantages of the {Matthews} correlation coefficient ({MCC}) over {F1} score and accuracy in binary classification evaluation},
  journal = {BMC Genomics},
  volume  = {21},
  number  = {1},
  pages   = {6},
  year    = {2020},
  doi     = {10.1186/s12864-019-6413-7}
}

@article{cooper1979,
  author  = {J. A. Cooper and R. Saracci and P. Cole},
  title   = {Describing the validity of carcinogen screening tests},
  journal = {British Journal of Cancer},
  volume  = {39},
  number  = {1},
  pages   = {87--89},
  year    = {1979},
  doi     = {10.1038/bjc.1979.10}
}

@article{kuncheva2003,
  author  = {Ludmila I. Kuncheva and Christopher J. Whitaker},
  title   = {Measures of Diversity in Classifier Ensembles and Their Relationship with the Ensemble Accuracy},
  journal = {Machine Learning},
  volume  = {51},
  number  = {2},
  pages   = {181--207},
  year    = {2003},
  doi     = {10.1023/A:1022859003006}
}

@inproceedings{marx2020,
  author    = {Charles Marx and Flavio Calmon and Berk Ustun},
  title     = {Predictive Multiplicity in Classification},
  booktitle = {Proceedings of the 37th International Conference on Machine Learning},
  series    = {Proceedings of Machine Learning Research},
  volume    = {119},
  pages     = {6765--6774},
  year      = {2020},
  eprint    = {1909.06677},
  archivePrefix = {arXiv}
}

@article{watsondaniels2023,
  author  = {Jamelle Watson-Daniels and David C. Parkes and Berk Ustun},
  title   = {Predictive Multiplicity in Probabilistic Classification},
  journal = {Proceedings of the AAAI Conference on Artificial Intelligence},
  volume  = {37},
  number  = {9},
  pages   = {10306--10314},
  year    = {2023},
  doi     = {10.1609/aaai.v37i9.26227}
}

@inproceedings{hsu2022,
  author    = {Hsiang Hsu and Flavio du Pin Calmon},
  title     = {Rashomon Capacity: A Metric for Predictive Multiplicity in Classification},
  booktitle = {Advances in Neural Information Processing Systems 35},
  pages     = {28988--29000},
  year      = {2022},
  doi       = {10.52202/068431-2101},
  eprint    = {2206.01295},
  archivePrefix = {arXiv}
}

@misc{somepalli2022,
  author        = {Gowthami Somepalli and Liam Fowl and Arpit Bansal and Ping Yeh-Chiang and Yehuda Dar and Richard Baraniuk and Micah Goldblum and Tom Goldstein},
  title         = {Can Neural Nets Learn the Same Model Twice? Investigating Reproducibility and Double Descent from the Decision Boundary Perspective},
  year          = {2022},
  eprint        = {2203.08124},
  archivePrefix = {arXiv},
  primaryClass  = {cs.LG},
  note          = {To appear in CVPR 2022}
}

@article{vidal2003,
  author  = {Guifr{\'e} Vidal},
  title   = {Efficient Classical Simulation of Slightly Entangled Quantum Computations},
  journal = {Physical Review Letters},
  volume  = {91},
  number  = {14},
  pages   = {147902},
  year    = {2003},
  doi     = {10.1103/PhysRevLett.91.147902},
  eprint  = {quant-ph/0301063},
  archivePrefix = {arXiv}
}

@article{cortes1995support,
  author  = {Corinna Cortes and Vladimir Vapnik},
  title   = {Support-vector networks},
  journal = {Machine Learning},
  volume  = {20},
  number  = {3},
  pages   = {273--297},
  year    = {1995},
  doi     = {10.1007/BF00994018}
}

@article{schuld2019quantum,
  author  = {Maria Schuld and Nathan Killoran},
  title   = {Quantum Machine Learning in Feature {Hilbert} Spaces},
  journal = {Physical Review Letters},
  volume  = {122},
  number  = {4},
  pages   = {040504},
  year    = {2019},
  doi     = {10.1103/PhysRevLett.122.040504}
}

@misc{lloyd2020quantum,
  author        = {Seth Lloyd and Maria Schuld and Aroosa Ijaz and Josh Izaac and Nathan Killoran},
  title         = {Quantum Embeddings for Machine Learning},
  year          = {2020},
  eprint        = {2001.03622},
  archivePrefix = {arXiv},
  primaryClass  = {quant-ph}
}

@article{RebentrostEtAl2014PRL,
  author  = {Patrick Rebentrost and Masoud Mohseni and Seth Lloyd},
  title   = {Quantum Support Vector Machine for Big Data Classification},
  journal = {Physical Review Letters},
  volume  = {113},
  number  = {13},
  pages   = {130503},
  year    = {2014},
  doi     = {10.1103/PhysRevLett.113.130503}
}

@article{HuangEtAl2021NatComm,
  author  = {Hsin-Yuan Huang and Michael Broughton and Masoud Mohseni and Ryan Babbush and Sergio Boixo and Hartmut Neven and Jarrod R. McClean},
  title   = {Power of Data in Quantum Machine Learning},
  journal = {Nature Communications},
  volume  = {12},
  pages   = {2631},
  year    = {2021},
  doi     = {10.1038/s41467-021-22539-9}
}

@article{Thanasilp2024,
  author  = {Supanut Thanasilp and Samson Wang and M. Cerezo and Zo{\"e} Holmes},
  title   = {Exponential Concentration in Quantum Kernel Methods},
  journal = {Nature Communications},
  volume  = {15},
  pages   = {5200},
  year    = {2024},
  doi     = {10.1038/s41467-024-49287-w},
  eprint  = {2208.11060},
  archivePrefix = {arXiv}
}

@article{Schnabel_2025,
  author  = {Jan Schnabel and Marco Roth},
  title   = {Quantum Kernel Methods under Scrutiny: A Benchmarking Study},
  journal = {Quantum Machine Intelligence},
  volume  = {7},
  number  = {1},
  pages   = {58},
  year    = {2025},
  doi     = {10.1007/s42484-025-00273-5},
  eprint  = {2409.04406},
  archivePrefix = {arXiv}
}

@misc{D_Amore_2026,
  author        = {Francesco D'Amore and Luca Mariani and Carlo Mastroianni and Francesco Plastina and Luca Salatino and Jacopo Settino and Andrea Vinci},
  title         = {Assessing Projected Quantum Kernels for the Classification of {IoT} Data},
  year          = {2026},
  eprint        = {2505.14593},
  archivePrefix = {arXiv},
  primaryClass  = {quant-ph}
}

@article{gonen2011multiple,
  author  = {Mehmet G{\"o}nen and Ethem Alpayd{\i}n},
  title   = {Multiple Kernel Learning Algorithms},
  journal = {Journal of Machine Learning Research},
  volume  = {12},
  pages   = {2211--2268},
  year    = {2011}
}

@misc{yyvr-dtsb,
  author        = {Vladyslav Bohun and Illia Lukin and Mykola Luhanko and Georgios Korpas and Philippe J. S. De Brouwer and Mykola Maksymenko and Maciej Koch-Janusz},
  title         = {Entanglement Scaling in Matrix Product State Representation of Smooth Functions and Their Shallow Quantum Circuit Approximations},
  year          = {2024},
  eprint        = {2412.05202},
  archivePrefix = {arXiv},
  primaryClass  = {quant-ph}
}

@article{truchon2007,
  author  = {Jean-Fran{\c{c}}ois Truchon and Christopher I. Bayly},
  title   = {Evaluating Virtual Screening Methods: Good and Bad Metrics for the ``Early Recognition'' Problem},
  journal = {Journal of Chemical Information and Modeling},
  volume  = {47},
  number  = {2},
  pages   = {488--508},
  year    = {2007},
  doi     = {10.1021/ci600426e}
}

\end{document}